\documentclass{aa}
\usepackage[varg]{txfonts}
\usepackage{bm}
\usepackage{mathtools}
\usepackage{amssymb}
\usepackage{graphicx}
\usepackage{epstopdf}
\usepackage{natbib}
\usepackage{color}

\bibpunct{(}{)}{;}{a}{}{,}
\usepackage{multirow}
\usepackage{booktabs}

\newlength{\verticalcompensationlength}
\newcounter{verticalcompensationrows}

\begin{document}

   \title{Periodic orbits in the 1:2:3 resonant chain and their impact on the orbital dynamics of the Kepler-51 planetary system}

   \author{Kyriaki I. Antoniadou$^1$    \and       George Voyatzis$^1$
          }

   \institute{$^1$Department of Physics, Aristotle University of Thessaloniki, 54124,
              Thessaloniki, Greece\\
                      \email{kyant@auth.gr}
         }

\titlerunning{Orbital stability of the three-planet system Kepler-51}

\authorrunning{K. I. Antoniadou and G. Voyatzis}

\date{Received XXXX; Accepted YYYY}

  \abstract
   {}
   {Space missions have discovered a large number of exoplanets evolving in (or close to) mean-motion resonances (MMRs) and resonant chains. Often, the published data exhibit very high uncertainties due to the observational limitations that introduce chaos into the evolution of the system on especially shorter or longer timescales. We propose a study of the dynamics of such systems by exploring particular regions in phase space.}
   {We exemplify our method by studying the long-term orbital stability of the three-planet system Kepler-51 and either favor or constrain its data. It is a dual process which breaks down in two steps: the computation of the families of periodic orbits in the 1:2:3 resonant chain and the visualization of the phase space through maps of dynamical stability.}
   {We present novel results for the general four-body problem. Stable periodic orbits were found only in the low-eccentricity regime. We demonstrate three possible scenarios safeguarding Kepler-51, each followed by constraints. Firstly, there are the 2/1 and 3/2 two-body MMRs, in which $e_b<0.02$, such that these two-body MMRs last for extended time spans. Secondly, there is the 1:2:3 three-body Laplace-like resonance, in which $e_c<0.016$ and $e_d<0.006$ are necessary for such a chain to be viable. Thirdly, there is the combination comprising the 1/1 secondary resonance inside the 2/1 MMR for the inner pair of planets and an apsidal difference oscillation for the outer pair of planets in which the observational eccentricities, $e_b$ and $e_c$, are favored as long as $e_d\approx 0$.}
   {With the aim to obtain an  optimum deduction of the orbital elements, this study showcases the need for dynamical analyses based on periodic orbits performed in parallel to the fitting processes.}

   \maketitle
%
%-------------------------------------------------------------------

\section{Introduction} \label{intro}
The $Kepler$, K2, and TESS missions have revealed 3673 planetary systems, and 814 of them feature multiple planets\footnote{As seen in \textit{exoplanet.eu} in March 2022} \citep[see e.g.,][]{lismass,fab14,peaspod}. Apart from various properties and architectures, such as the number of planets, the planetary period, and radii ratios observed among them, planetary pairs and triplets seem to be evolving near (just outside) or in two-body mean-motion resonances (MMRs) and three-body Laplace-like resonances (resonant chains), respectively. In particular, five out of six planets of TOI-178 are in the 2:4:6:9:12 resonant chain \citep{leleu21}, four planets of the HR 8799 are in the 8:4:2:1 chain \citep{hr8799}, four planets of the system K2-32 evolve near the 1:2:5:7 resonant chain \citep{k232}, four planets of the system Kepler-223 are in a 3:4:6:8 resonant chain \citep{kep223}, while five planets of the K2-138 are in the 3/2 MMR \citep{k2138_18}. Other examples include the multiple-planet systems Trappist-1 \citep{gillon17}, Kepler-18 \citep{kep18}, Kepler-30 \citep{kep30}, Kepler-47 \citep{kep47}, Kepler-60 \citep{kep60}, and Kepler-82 \citep{kepler82}. 

Recent studies have been devoted to the formation history and dynamical evolution of multiple-planet systems that yield exoplanets locked in or close to a resonant chain \citep[e.g.,][]{chainsOli20,pichmorby20}. \citet{lisgav21} numerically analyzed the long-term orbital stability of three-planet systems where the planetary longitudes and orbital separation vary, while \citet{Siegel_2021} performed a numerical exploration of equillibria dependent on the libration of the three-body resonant angles for chains consisting of first-order MMRs. Furthermore, \citet{spock} provided a stability estimator that distinguishes quickly the unstable from the stable planetary configurations among estimated orbital parameters and masses of multiple-planet systems. \citet{charalbeau18} chose different planetary masses and illustrated maps  with stable and unstable domains of three-planet systems at which the mean-motion ratios of the planets were altered, while \citet{petit21} analytically approximated first-order resonant chains of three massive planet systems with an averaged Hamiltonian. 

Planetary masses and eccentricities can be precisely extracted by the transit-timing variation (TTV) method for exoplanets close to an MMR. However, employing numerical analyses or $N$-body simulation algorithms in order to reject orbital parameters that lead to instability events (and which cannot be estimated by observations) may become computationally expensive and time-consuming. Additionally, the published observational data often possess very large deviations that render the system chaotic in nature. Instead, by spotting the stable periodic orbits close to the exoplanetary system (for a given MMR or resonant chain, planetary masses and eccentricity values) can immediately unravel the regions where the stability is guaranteed for long-time spans. Hence, the boundaries for all the orbital elements, even the longitudes of pericenter and mean anomalies, can be delineated as the dynamical phase space of resonant exoplanets is crystallized. 

\citet{hadjmich81} computed the first stable periodic orbits in the planar General 4-Body Problem (G4BP) \citep[see e.g.][]{hadjNbody,michgri89} for the Galilean satellites of Jupiter in the 1:2:4 resonant chain. Following their work, we apply this methodology to the exoplanets. Here, we focus on the system Kepler-51 and aim to provide hints and constraints that favor its survival. We study the dynamics and long-term stability of the system via the families of periodic orbits in the planar G4BP and extend our methodology for pairs of resonant massive exoplanets in mutually inclined orbits \citep{av12,av14} and coplanar orbits \citep{a16,av16} to the case of triplets of massive coplanar exoplanets in resonant chains (G4BP).

Our work is organized as follows. In Sect.~\ref{main}, we provide the key points and tools employed to obtain our dynamical analysis. In Sect. \ref{dynKep}, we discuss possible regions of long-term stable evolution in the dynamical vicinity of Kepler-51 based on the periodic orbits in the G4BP and conclude in Sect. \ref{fin}. In Appendix \ref{model}, we provide the equations of motion for three-planet systems (\ref{eqsm}), define the symmetric periodic orbits (\ref{pc}), and elaborate on the continuation method followed for the 1:2:3 resonant periodic orbits (\ref{ApFams}). In Appendix \ref{stab}, we provide details and examples on the linear stability of the periodic orbits in the G4BP (\ref{linear}) and the chaotic indicator used (\ref{chaos}).

\section{Main aspects of the methodology}\label{main}

We considered three massive planets revolving around a star on coplanar orbits. When viewed in an inertial frame of reference\footnote{In Appendix \ref{eqsm}, we provide more details about the model set-up and the inertial and rotating frames of reference.}, these orbits correspond to Keplerian ellipses with heliocentric osculating elements, namely, the semimajor axes, $a_i$, the eccentricities, $e_i$, the longitudes of pericenter, $\varpi_i$, and the mean anomalies, $M_i$ $(i=1,2,3)$, where subscript 1 refers to the inner planet. In computations, we use the normalized set of units where the semimajor axis of the inner planet is equal to 1, $G\times m_{\rm total}=1$, and, subsequently, the period of the inner planet is equal to about $2\pi$. 

We chose the relative frequencies, $\omega_i$ $(i=1,2,3)$, of the three planets of period $T_i=\frac{2\pi}{\omega_i}$ $(i=1,2,3)$ in the inertial frame, so that they be commensurable as follows:
\begin{equation}
\frac{\omega_2-\omega_1}{\omega_3-\omega_1}\approx\frac{P}{Q},
\end{equation}
where $P,Q\in \mathbb{Z}^*$.

Then, in the rotating frame of reference we can define the relative period, $T$, of a system where planet 2 with period $T_2$ revolves $P$ times about planet 1 with period $T_1$, while planet 3 with period $T_3$ revolves $Q$ times around planet 1, that is:
\begin{equation}\label{PQ}
T\approx\frac{T_1}{1-\frac{T_1}{T_2}}\;P\approx\frac{T_1}{1-\frac{T_1}{T_3}}\;Q.
\end{equation}

Certain orbits that fulfill a specific periodicity condition in the rotating frame are called symmetric periodic orbits (see Appendix \ref{pc} for more details). These periodic orbits correspond to the exact location of the MMR. In this work, we focus exclusively on the symmetric elliptic (resonant) periodic orbits, which have $\varpi_i=0$ or $\pi$ $(i=1,2,3)$\footnote{Asymmetric periodic orbits also exist, having longitudes of pericenter different than 0 or $\pi$, which are not considered herein.}.

Described within another context, the periodic orbits are the fixed (or periodic) points of a Poincar\'e map and correspond also to the fixed points (or stationary solutions) of an averaged Hamiltonian, as long as the latter is accurate enough. Following the continuation schemes explained in Appendix \ref{pc} and \ref{ApFams}, these points form the so-called characteristic curves or families of periodic orbits. 

The periodic orbits depend on the resonant angles, $\theta_i$ $(i=1,...,4)$, which, given the 1:2:3 resonant chain studied, take the following form:
\begin{equation}\label{eqangles}
\begin{aligned}
\theta_1&=2\lambda_2-\lambda_1-\varpi_1,\\
\theta_2&=2\lambda_2-\lambda_1-\varpi_2,\\
\theta_3&=3\lambda_3-2\lambda_2-\varpi_3,\\
\theta_4&=3\lambda_3-2\lambda_2-\varpi_2,\\
\phi_L&=\theta_4-\theta_2=3\lambda_3-4\lambda_2+\lambda_1,
\end{aligned}
\end{equation}
with $\lambda_i=\varpi_i+M_i$ as the mean longitude, $\Delta\varpi_{21}=\varpi_2-\varpi_1$ and $\Delta\varpi_{32}=\varpi_3-\varpi_2$ as the apsidal differences, and with $\phi_L$ as the Laplace angle. 

If the periodic orbit is symmetric, the angles in Eq. \ref{eqangles} are equal to either 0 or $\pi$. In order to distinguish the families of periodic orbits that belong to different configurations, we present them on the $(e_1\cos\theta_1,e_2\cos\theta_2)$ and $(e_2\cos\theta_2,e_3\cos\theta_3)$ planes.

The characteristic curves of the families change significantly, when the mass-ratio changes \citep[see e.g.][]{av12} for the 2/1 MMR and \citep[see e.g.][]{av14} for the 3/2, 5/2, 3/1 and 4/1 MMRs. If the mass values change, but still retain the same order of magnitude and preserve their ratio, then the main properties of the families (location and stability) do not show significant variations \citep[see e.g.][]{voyatzis08}. Therefore, in order to distinguish the families that belong to the same configuration but have been computed for different mass-ratios, we introduce two planetary mass-ratios; one for the innermost and one for the outermost pair, namely, $\rho_i=\frac{m_c}{m_b}$ and $\rho_o=\frac{m_d}{m_c}$, respectively.

It is widely known in Hamiltonian systems that stable periodic orbits are surrounded by invariant tori where the motion is regular and quasi-periodic, whereas the neighborhood of unstable periodic orbits can give rise to instability events (either weak chaos or strong chaos with collisions or escapes). In the neighborhood of stable periodic orbits, all the resonant angles librate, while all or some of them exhibit rotation in the vicinity of the unstable ones.  

\begin{table*}[h!]
\centering
\caption{Published data related to the planetary masses of Kepler-51 considered in this study.}
%\scalebox{0.9}{
\begin{tabular}[b]{lcccccc}
\toprule
$m_S\left(M_{\odot}\right)$&$m_b\left(M_\oplus\right)$&$m_c\left(M_\oplus\right)$&$m_d\left(M_\oplus\right)$&$\rho_i=\frac{m_c}{m_b}$&$\rho_o=\frac{m_d}{m_c}$&References\\
\cmidrule{1-7}
$1.04$          &$2.1$    &$4.0$                &$7.6$          &1.91&1.90&1\\
\cmidrule{1-7}
$0.985$ &$3.69$   &$4.43$   &$5.70$     &1.20&1.29&2\\
\cmidrule{1-7}
$1.053$ &$1.1^{a}/2.3^{b}$    &$2.8^{a}/3.4^{b}$        &$4.4^{a}/5.2^{b}$      &$2.54^{a}/1.48^{b}$&$1.57^{a}/1.53^{b}$&3\\
\bottomrule
\end{tabular}

\tablefoot{Both $^{(a)}$default and $^{(b)}$high-mass priors considered in (3).}
\tablebib{(1) \citet{masuda}; (2) \citet{libby20}; (3) \citet{bat21}}\label{tmasses}
\end{table*}

\begin{table*}[h!]
\centering
\caption{Published data that were utilized for the constraints on Kepler-51$^{(1)}$.}
\begin{tabular}[b]{lccc}
\toprule
Parameter&Kepler-51b&Kepler-51c&Kepler-51d\\
\cmidrule{1-4}
$m_p\left(M_\oplus\right)$     &$2.1^{+1.5}_{-0.8}$    &$4.0 \pm 0.4$                &$7.6 \pm 1.1$\\
\cmidrule{2-4}
$T\left(days\right)$           &$45.1540\pm 0.0002$    &$85.312^{+0.003}_{-0.002}$   &$130.194^{+0.005}_{-0.002}$\\
\cmidrule{2-4}
$a\left(au\right)$             &$0.2514\pm 0.0097$     &$0.384\pm 0.015$             &$0.509\pm 0.020$\\
\cmidrule{2-4}
$e$                            &$0.04\pm 0.01$         &$0.014^{+0.013}_{-0.009}$    &$0.008^{+0.011}_{-0.008}$\\
\cmidrule{2-4}
$T_0\left(BJD-2454833\right)$  &$881.5977\pm 0.0004$   &$892.509\pm 0.003$           &$862.9323\pm 0.0004$\\
\bottomrule
\end{tabular}
\tablefoot{Observational values provided by $^{(1)}$\citet{masuda}.}\label{tab}
\end{table*}

The libration of all the resonant angles $\theta_i$ $(i=1,...,4)$ signifies a two-body MMR (hereafter denoted as $R_L$), where $\frac{n_2}{n_1}=(\frac{a_1}{a_2})^{-3/2}\approx \frac{p_1+q_1}{p_1}$ and $\frac{n_3}{n_2}=(\frac{a_2}{a_3})^{-3/2}\approx \frac{p_2+q_2}{p_2}$ with $p_i,q_i\in \mathbb{Z}^*$ and $q_i$ being the order of the MMR, implying that the libration of the Laplace angle, $\phi_L$, is simply the consequence of the two independent MMRs which overlap. However, $\phi_L$ can librate while the rest of the resonant angles, $\theta_i$, rotate. The latter indicates a state where the pairs of planets are not locked in a two-body MMR, but the triplet is locked in a three-body resonance (resonant chain, denoted as $R_T$), where $(\frac{n_2}{n_3})^{-1}\approx \frac{p+q-s}{q}-\frac{p}{q}(\frac{n_1}{n_2})$ with $p,q,s \in \mathbb{Z}^*$ and $s$ being the order of the resonance \citep[see e.g.,][]{modern}. These resonance mechanisms provide a phase protection. Inside an MMR, we can observe a secondary resonance (denoted as $R_S$) %\citep[see e.g.][]{secondary}, 
or an apsidal difference oscillation (denoted as $R_A$). In the former case, we consider the frequency ratio between the libration frequency of $\theta_1$ and the rotation frequency of $\theta_2$ \citep[see e.g.,][]{secondary}, while in the latter case, only the apsidal difference oscillates and the rest of the angles rotate. At the non-resonant low-eccentricity orbits, an apsidal resonance, where only the apsidal difference librates, may become apparent. In the following, we also use a notation of $R$ with two subscripts, namely $R_{i,o}$, in order to denote two separate behaviors of the inner and outer pairs of planets. For instance, the combination of a secondary resonance of the inner pair with either an MMR or apsidal difference oscillation of the outer pair is denoted as $R_{S,L}$ or $R_{S,A}$, respectively. The above dynamical mechanisms have been extensively showcased in the neighborhood of periodic orbits by \citet[][for the minor-body dynamics]{spis} and are all observed in the regular domains in phase space, where even massive planets can evolve in a stable way \citep{av16}.

We study the long-term orbital stability of a three-planet systems based on two methods: i) linear stability of the periodic orbits (explained in Appendix \ref{linear}) and ii) maps of dynamical stability (DS-maps) by computing the chaotic indicator DFLI (see Appendix \ref{chaos}). The first method yields either stable or unstable periodic orbits, which are either colored blue or red along the families. Based on the stable periodic orbits, we seed the second method and construct DS-maps around the (near)-resonant exoplanets and provide a visualization of the boundaries of the orbital elements securing regular motion.

\section{Dynamical constraints on Kepler-51}\label{dynKep}

\begin{figure*}[ht!]%[ht!]
%\plottwo{123e1e2_all.eps}{123e2e3_all.eps}
\includegraphics[width=0.99\textwidth]{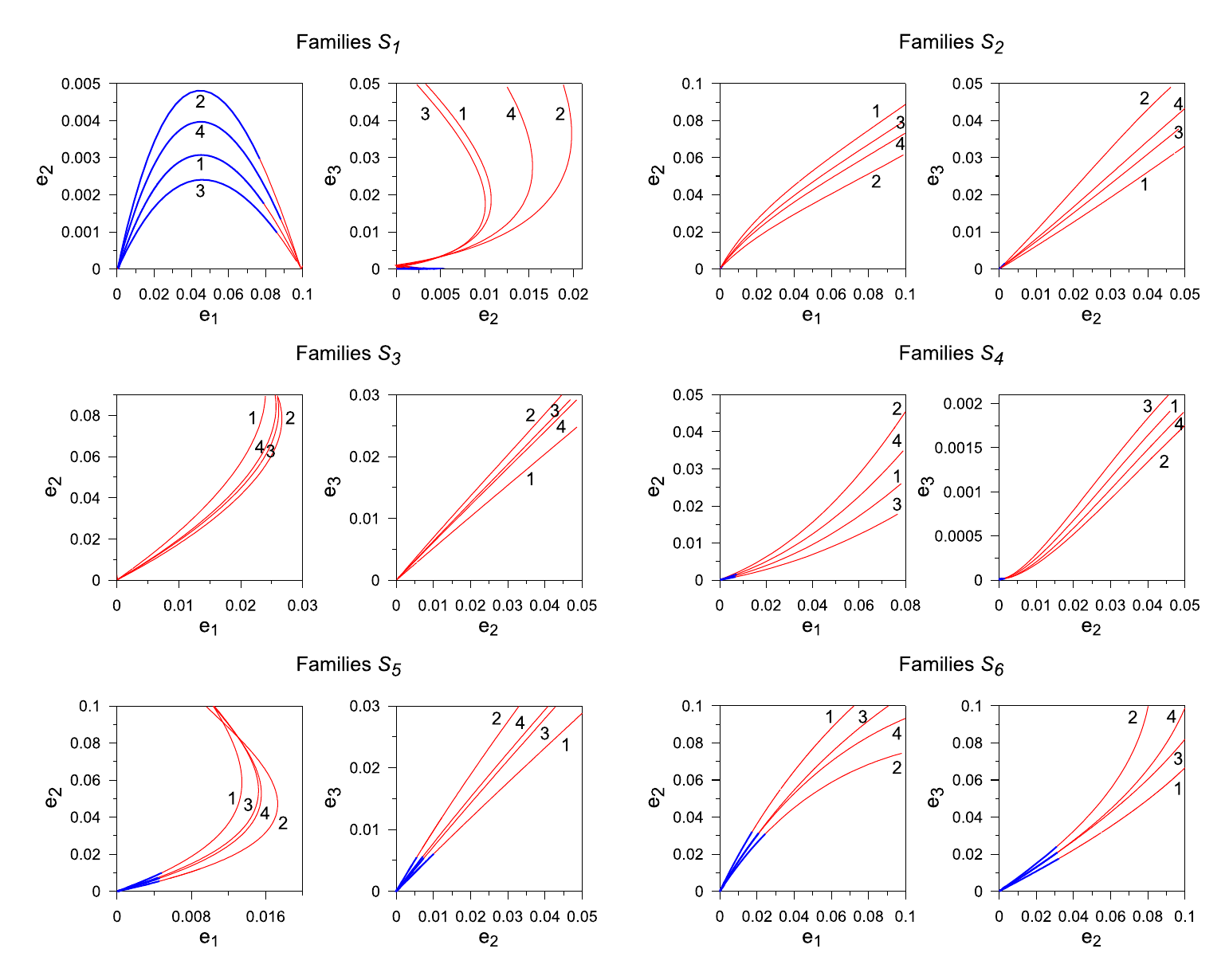}%\vspace{-0.5cm}\\
\caption{Groups of families, $S_i$ $(i=1...6)$, of planar symmetric periodic orbits in the 1:2:3 resonant chain with stable segments close to the dynamical vicinity of Kepler-51 presented on the $(e_1,e_2)$ and $(e_2,e_3)$ planes. Each of these six groups consists of families of the same configuration (shown in Table \ref{tconf}), but with different planetary mass-ratios, that is, $\rho_i=\frac{m_c}{m_b}$ and $\rho_o=\frac{m_d}{m_c}$ (shown in Table \ref{tmasses} and labeled from 1 to 4 on each panel). Label 1 corresponds to \citet{masuda}, label 2 to \citet{libby20}, label 3 to default, and label 4 to high-mass priors considered by \citet{bat21}. The stable (unstable) periodic orbits are depicted by blue (red) solid curves.}\label{fams_all_masses}
\end{figure*}

The system Kepler-51 has been regularly studied  \citep[see e.g.,][]{steffen13,had14,masuda,mor16,holcz16,had17,berg18,gajdos,libby20,bat21}.  Its innermost planetary pair evolves close to the 2/1 MMR, according to the period ratio $\frac{T_c}{T_b}=1.89$, while the outermost one is in the 3/2 MMR, as $\frac{T_d}{T_c}=1.53$. In what follows, we present the families of periodic orbits in the 1:2:3 resonant chain and construct DS-maps that unveil the dynamical vicinity of Kepler-51 in order to explore the dynamical mechanisms that may guarantee its orbital stability.

\begin{figure*}[ht!]%[ht!]
$\begin{array}{c}
\includegraphics[width=0.99\textwidth]{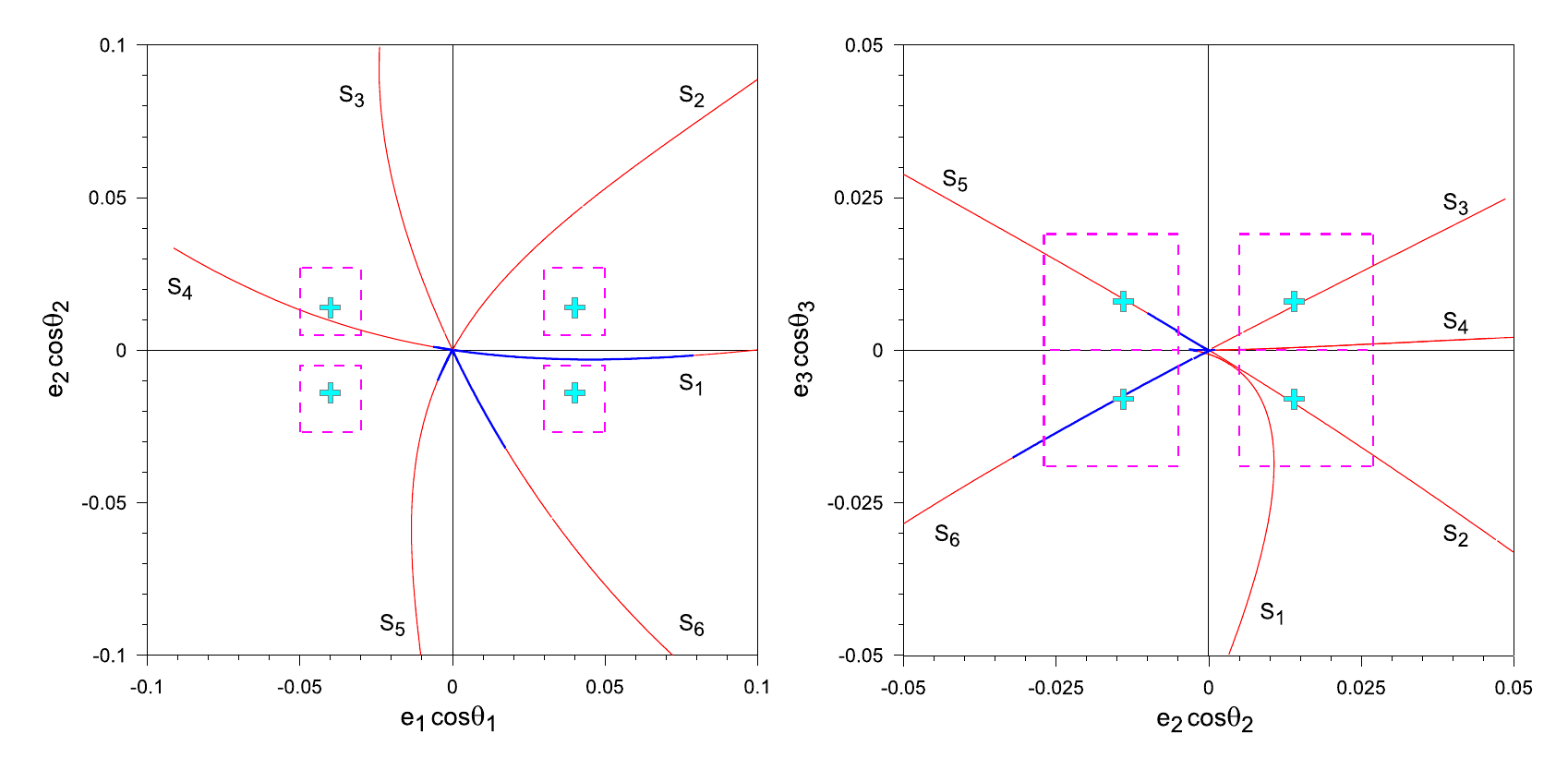}\\
\includegraphics[width=0.99\textwidth]{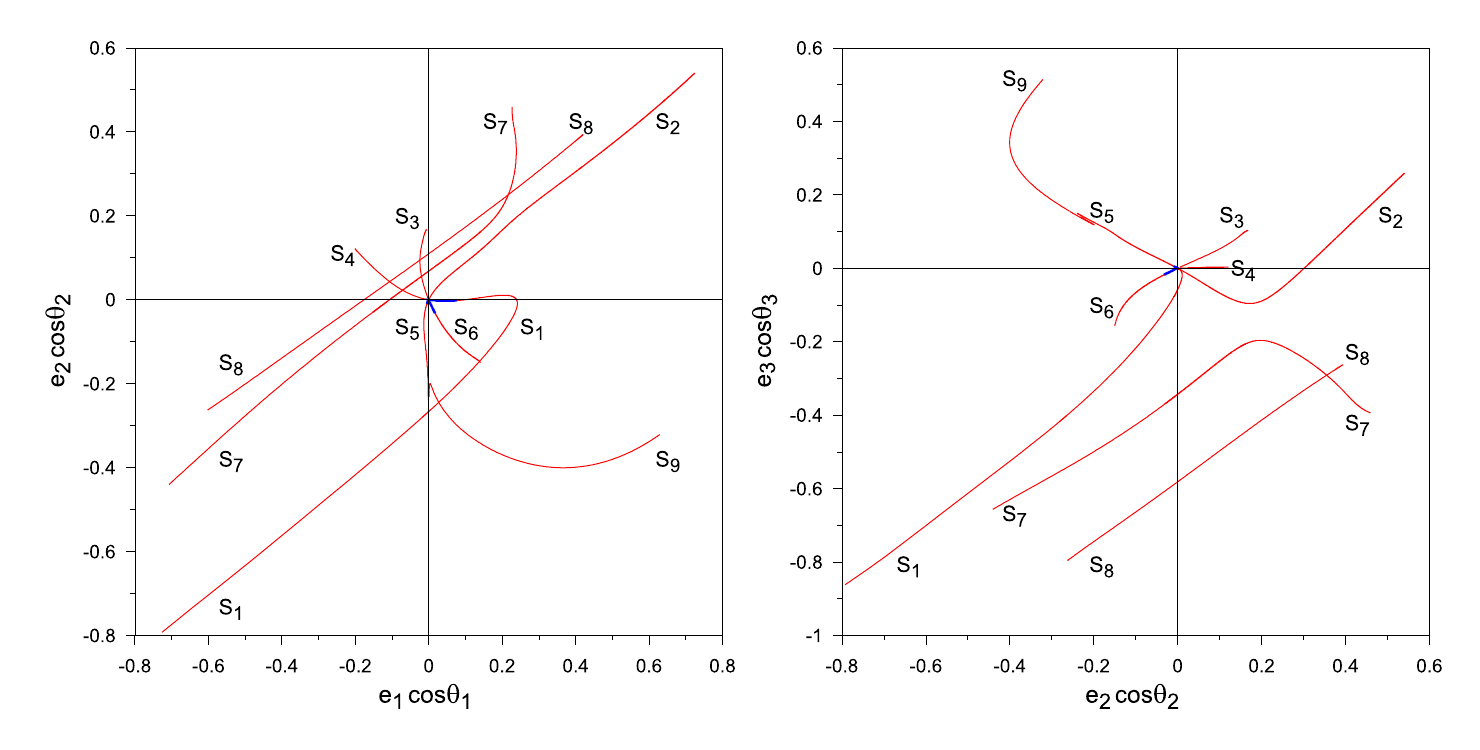}\\\end{array}$
\caption{Families, $S_i$ $(i=1...9)$, of planar symmetric periodic orbits in the 1:2:3 resonant chain close to the dynamical vicinity of Kepler-51 (top panels) and until their termination points (bottom panels) presented on the $(e_1\cos\theta_1,e_2\cos\theta_2)$ and $(e_2\cos\theta_2,e_3\cos\theta_3)$ planes. The angles for the different configurations of the families are shown in Table \ref{tconf}. Kepler-51 is identified in the top panels by the cyan "$+$" symbol, while the errors provided by \citet{masuda} are delineated by magenta dashed lines.}\label{123fams_z}
\end{figure*}

\begin{table*}[h!]
\centering
\caption{Configurations of families of symmetric periodic orbits in the 1:2:3 resonant chain.}\label{tconf}
\begin{tabular}[b]{lcccccccccccccc}
\toprule
Family &$\varpi_1$ & $\varpi_2$ & $\varpi_3$ & $M_1$ & $M_2$ & $M_3$ & $\theta_1$ & $\theta_2$ & $\theta_3$ & $\theta_4$ & $\phi_L$ & $e_1^*$ & $e_2^*$ & $e_3^*$\\
\cmidrule{1-15}
$S_1$   &0    &$\pi$   &0    &0    &$\pi$    &0    &0    &$\pi$    &0    &$\pi$    &0 & & &\\
        &     &   &    &    &   &    &    &   &    &    & &0.0647425 &0.0026077 &0.0\\
        &0    &$\pi$   &$\pi$  &0    &$\pi$    &$\pi$  &0    &$\pi$    &$\pi$  &$\pi$    &0 & & &\\
                                &     &  & &    &  &  &    &    &  &    & &0.0991335 &0.0 &0.0006738\\
                                &0    &0     &$\pi$  &0    &0      &$\pi$  &0    &0      &$\pi$  &0      &0 & & &\\
                                &     &     &  &    &      &  &    &      &  &      & &0.2391378 &0.0 &0.0581071\\
                                &0    &$\pi$   &$\pi$  &0    &$\pi$    &$\pi$  &0    &$\pi$    &$\pi$  &$\pi$    &0 &&& \\
                                &     &   &  &    &    &  &    &    &  &    & &0.0&0.2666913&0.4032193 \\
                                &$\pi$  &$\pi$   &$\pi$  &$\pi$  &$\pi$    &$\pi$  &$\pi$  &$\pi$    &$\pi$  &$\pi$    &0 &&& \\
\cmidrule{1-15}
$S_2$   &0    &0     &$\pi$  &0    &0      &$\pi$  &0    &0      &$\pi$  &0      &0 &&&\\
        &     &     &  &    &      &  &    &      &  &      & &0.3948695&0.3015711&0.0\\
        &0    &0     &0    &0    &0      &0    &0    &0      &0    &0      &0 &&&\\
\cmidrule{1-15}
$S_3$     &$\pi$  &0     &$\pi$  &$\pi$  &0      &0    &$\pi$  &0      &0    &$\pi$    &$\pi$ &&&\\
\cmidrule{1-15}
$S_4$     &$\pi$  &0     &0    &$\pi$  &0      &0    &$\pi$  &0      &0    &0      &0 &&&\\
\cmidrule{1-15}
$S_5$     &$\pi$  &$\pi$   &0    &$\pi$  &$\pi$    &0    &$\pi$  &$\pi$    &0    &$\pi$    &0 &&& \\
\cmidrule{1-15}
$S_6$     &0    &$\pi$   &0    &0    &0      &$\pi$  &0    &$\pi$    &$\pi$  &0      &$\pi$ &&&\\
\cmidrule{1-15}
$S_7$   &0    &0     &$\pi$  &0    &0      &$\pi$  &0    &0      &$\pi$  &0      &0 &&&\\
        &    &     &  &    &      &  &    &      &  &      & &0.0&0.0682943&0.2815644\\
        &$\pi$  &0     &$\pi$  &$\pi$  &0      &$\pi$  &$\pi$  &0      &$\pi$  &0      &0 &&&\\
                                &  &     &  &  &      &  &  &      &  &      & &0.1066329&0.0&0.343046\\
                                &$\pi$  &$\pi$   &$\pi$  &$\pi$  &$\pi$    &$\pi$  &$\pi$  &$\pi$    &$\pi$  &$\pi$    &0 &&&\\
\cmidrule{1-15}
$S_8$   &0    &0     &$\pi$  &0    &0      &$\pi$  &0    &0      &$\pi$  &0      &0 &&&\\
                                &    &     &  &    &      &  &   &      &  &      & &0.0&0.1101254&0.488658\\
                                &$\pi$  &0     &$\pi$  &$\pi$  &0      &$\pi$  &$\pi$  &0      &$\pi$  &0      &0 &&&\\
                                &  &     &  & &      &  &  &      &  &      & &0.1764644&0.0&0.5815609\\
                                &$\pi$  &$\pi$   &$\pi$  &$\pi$  &$\pi$    &$\pi$  &$\pi$  &$\pi$    &$\pi$  &$\pi$    &0 &&&\\
\cmidrule{1-15}
$S_9$     &0    &$\pi$   &0    &0    &$\pi$    &0    &0    &$\pi$    &0    &$\pi$    &0 &&&\\
\bottomrule
\end{tabular}
\tablefoot{Transitions to different configurations as the families, $S_i$ $(i=1...9)$, of symmetric periodic orbits in the 1:2:3 resonant chain (presented in Fig. \ref{123fams_z}) evolve from the origin of the axes (circular family) up to high eccentricity values. The transitions along the families $S_7$ and $S_8$ begin from the configuration where $e_1>0$ and are monitored as they evolve thereafter. The eccentricity values $e_i^*$ ($i=1,2,3$) represent the periodic orbits where a change of a configuration takes place along the family.}
\end{table*}

\subsection{Families of symmetric periodic orbits for Kepler-51}\label{fams51}

Following the continuation method illustrated in Appendix~\ref{pc}, we set $P=3$ and $Q=4$ in Eq.~\ref{PQchain}, which for $k=4,$ yield the following:
\begin{equation}
\begin{aligned}
(\frac{T_2}{T_1}, \frac{T_3}{T_1}, \frac{T_3}{T_2})=(\frac{2}{1},\frac{3}{1},\frac{3}{2})
\end{aligned}
,\end{equation}
 namely, the resonant chain of the 2/1 and 3/2 MMRs is established for the innermost and outermost pairs of Kepler-51.

We then take into account the planetary eccentricity values of Kepler-51 (shown in Table \ref{tab}). In Fig. \ref{fams_all_masses}, we present the six groups of families of planar symmetric periodic orbits, $S_i$ $(i=1...6)$, in the 1:2:3 resonant chain, which reside very close to these observational values, on the $(e_1,e_2)$ and $(e_2,e_3)$ planes. Each of these six groups consists of families of the same configuration (shown in Table \ref{tconf}), but has different planetary mass-ratios, namely, $\rho_i=\frac{m_c}{m_b}$ and $\rho_o=\frac{m_d}{m_c}$ (shown in Table \ref{tmasses}). Each planetary-mass ratio is identified by a number on each panel. More particularly, label 1 corresponds to \citet{masuda}, label 2 to \citet{libby20}, label 3 to default, and label 4 to high-mass priors considered by \citet{bat21}.

We observe that the segments of stable periodic orbits are not altered in their extent since the considered planetary mass-ratios vary, even though the reflection of the four different sets of planetary masses (shown in Table \ref{tmasses}) on the the mass-ratios seems quite important and widely ranging; for instance, from 1.2 \citep{libby20} to 2.57 \citep{bat21}. When the dynamics is being studied, apart from the masses themselves, one major factor that affects it is the order of the mass values. More precisely, the divergence of the families and their stability segments as the mass-ratios vary are affected differently for giant Jovian planets and terrestrial ones; and all studies performed for the super-puff planets of Kepler-51 attribute masses of the same order $(10^{-6}-10^{-5})$.

Since the change on the dynamics is not significant among the various planetary masses considered in Table \ref{tmasses}, in what follows, we chose to focus on the study of \citet{masuda}. Our choice is also justified by both \citet{bat21} and \citet{libby20} who conclude that their mass values are less constrained and not improved when compared to \citet{masuda}.

In the top panels of Fig. \ref{123fams_z}, six families of planar periodic orbits, $S_i$ $(i=1...6)$, in the 1:2:3 resonant chain are presented on the signed-eccentricity planes. In the bottom panels, we provide all (nine) families up to high-eccentricity values, even though only low-eccentricity stable periodic orbits were found. The families were computed for the planetary masses of Kepler-51, shown in Table \ref{tab}. In Table \ref{tconf}, we provide the longitudes of pericenter, mean anomalies and resonant angles along each family together with the eccentricity values at which a transition of the configuration takes place. Given the observational values of the eccentricities (see Table \ref{tab}), depicted by a cyan "$+$" symbol, along with their deviations (shown as magenta dashed lines), we focus on the low-eccentricity regime (top panels). 

More precisely, solely the families $S_1$, $S_4$, $S_5$ and $S_6$ possess stable (blue colored) periodic orbits. These are found in the configurations $(\theta_1,\theta_2,\theta_3,\theta_4)$=$(0,\pi,0,\pi)$ and $(0,\pi,\pi,\pi)$ for the  $S_1$ family; $(\pi,0,0,0)$ for the $S_4 $ family; $(\pi,\pi,0,\pi)$ for the $S_5 $ family; and $(0,\pi,\pi,0)$ for the $S_6$ family.

\subsection{DS-maps and constraints for Kepler-51}\label{constraints}

Discerning distinct regions of regular and chaotic motion in phase space can sometimes be demanding for Hamiltonian systems of 5 degrees of freedom (planar G4BP). The use of chaotic indicators can assist in deciphering such domains. However, all indicators have their pros and cons \citep[see e.g.,][for a comparison between the Lyapunov Indicator (LI), the Mean Exponential Growth factor of Nearby Orbits (MEGNO), the Smaller Alignment Index (SALI), the FLI, and the Relative Lyapunov Indicator (RLI)]{maffione2011}. In this paper, we use a version of the FLI (called DFLI; see Appendix \ref{chaos} for its definition). Concerning dynamical systems such as the planetary ones, the DFLI was established as efficient and reliable by \citet{voyatzis08}.

\begin{figure*}[tp!]%[ht!]
$\begin{array}{cc}
\includegraphics[width=0.48\textwidth]{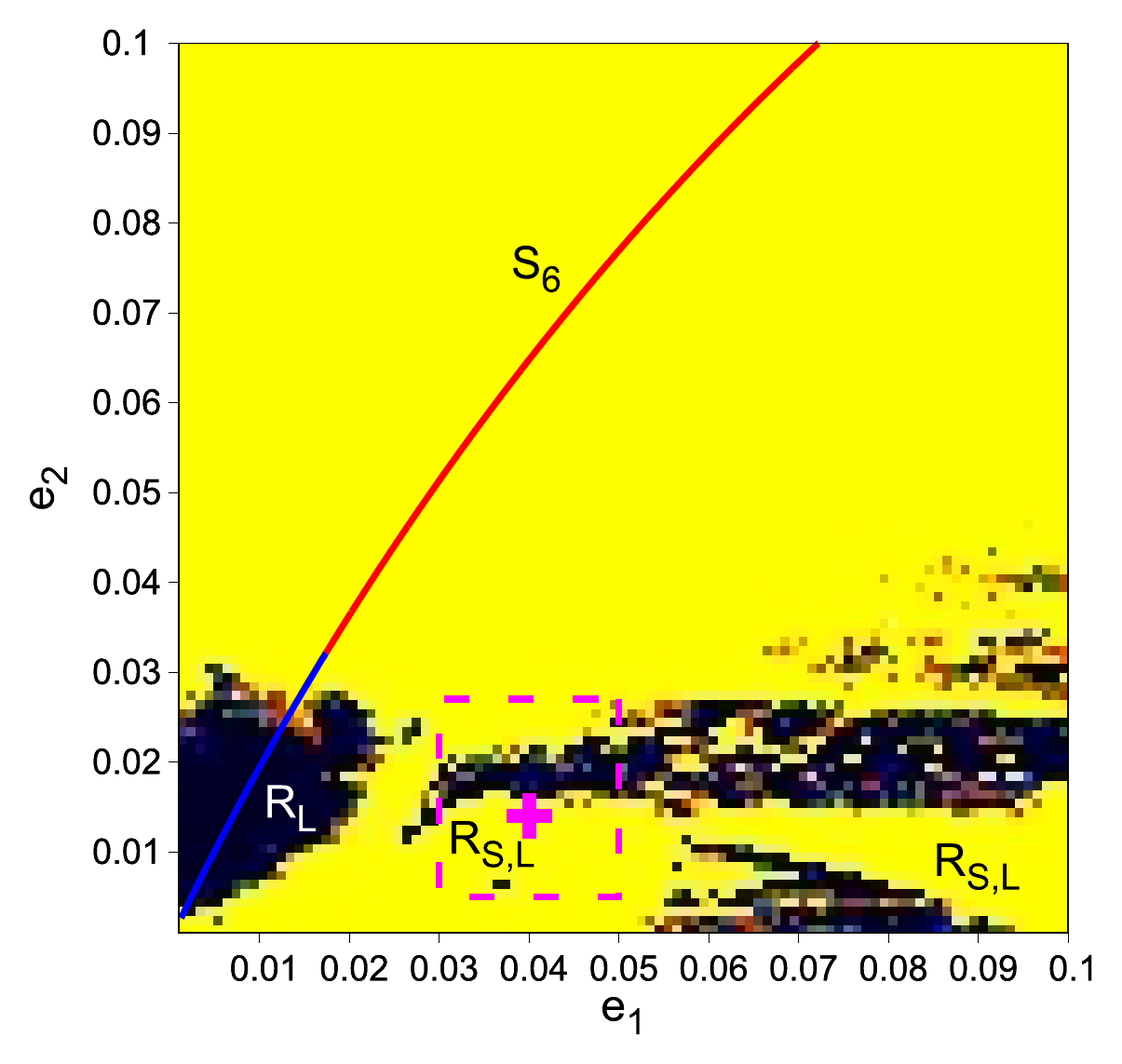}&\includegraphics[width=0.48\textwidth]{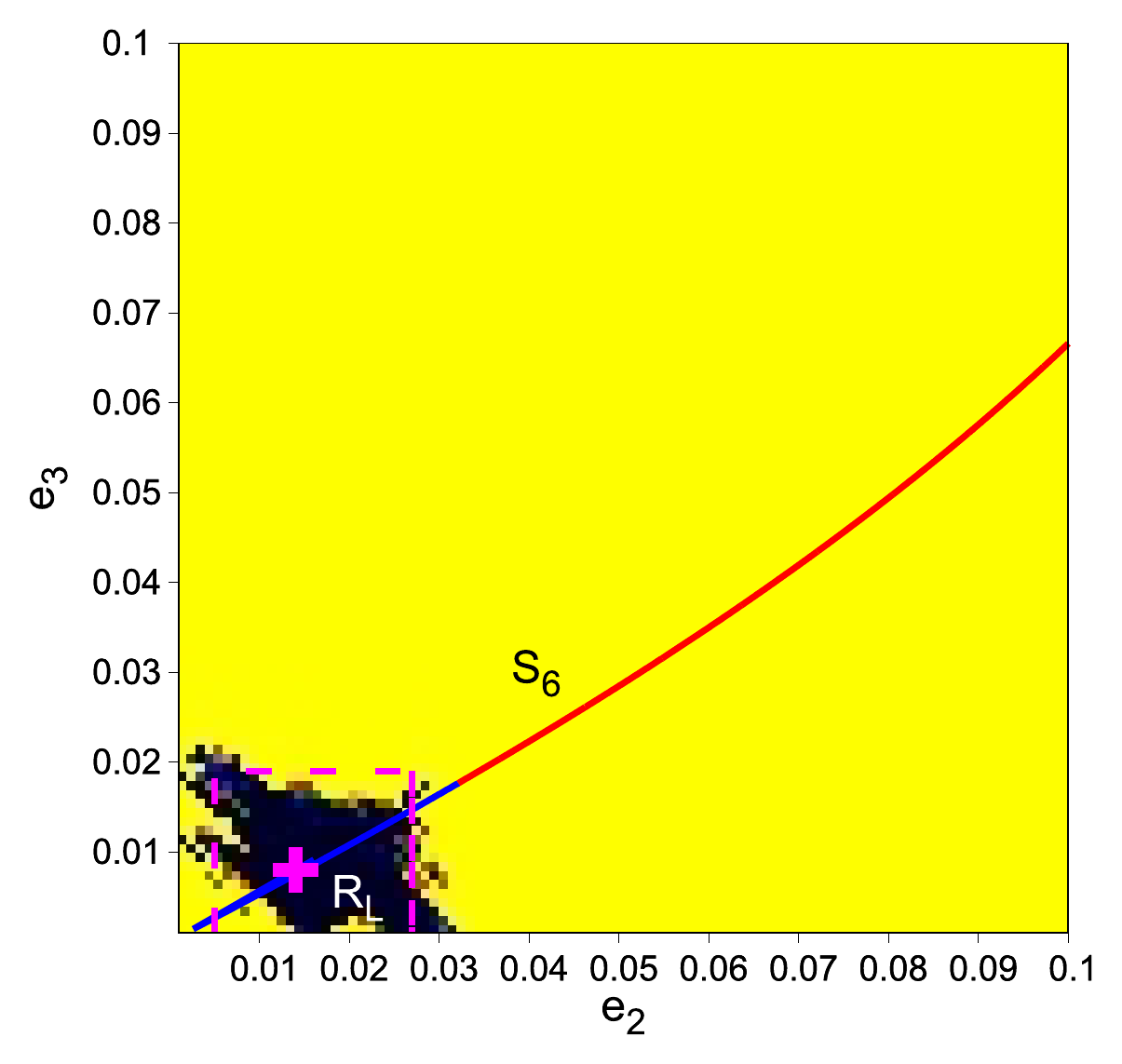}\\
\includegraphics[width=0.48\textwidth]{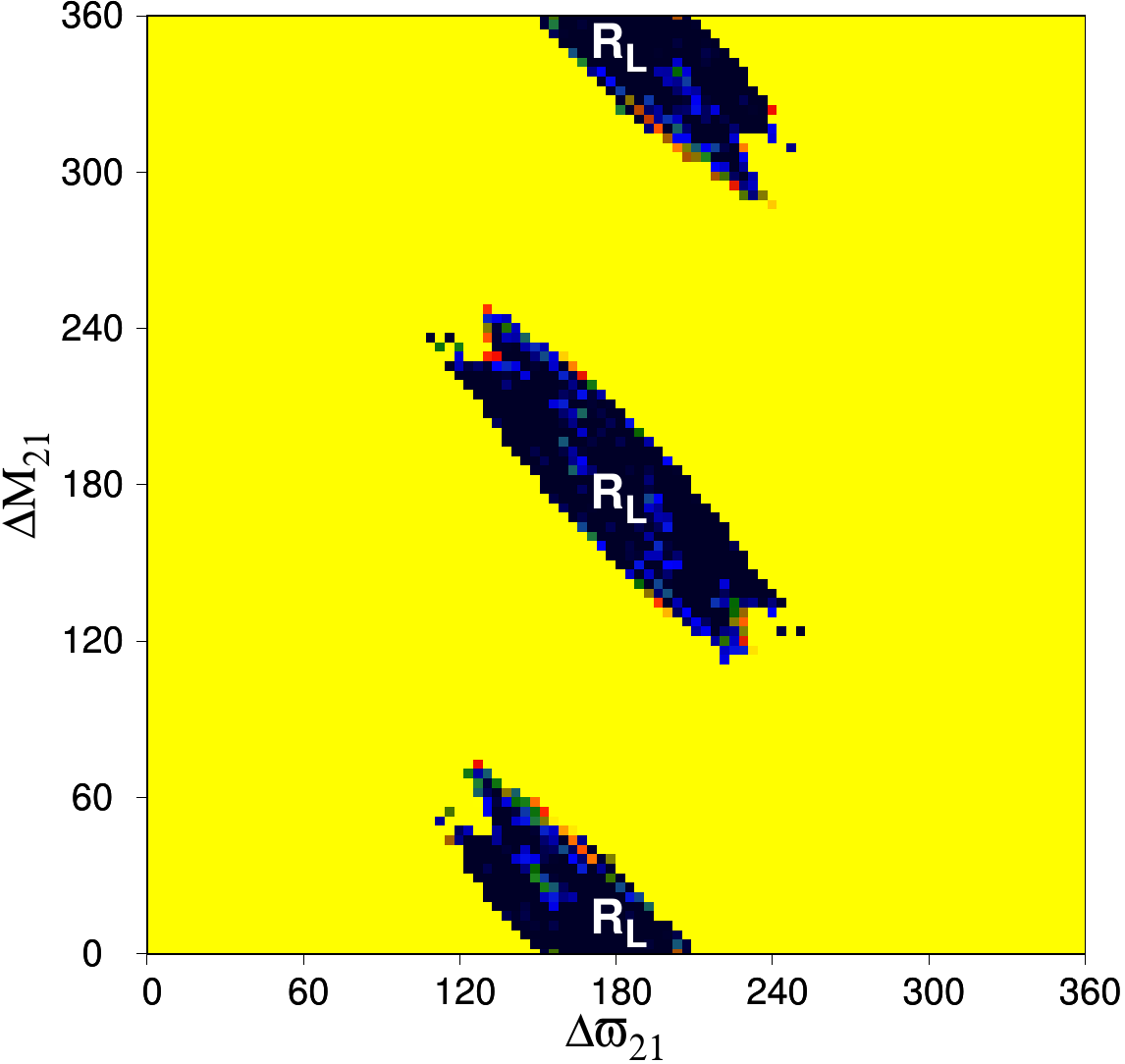}&\includegraphics[width=0.48\textwidth]{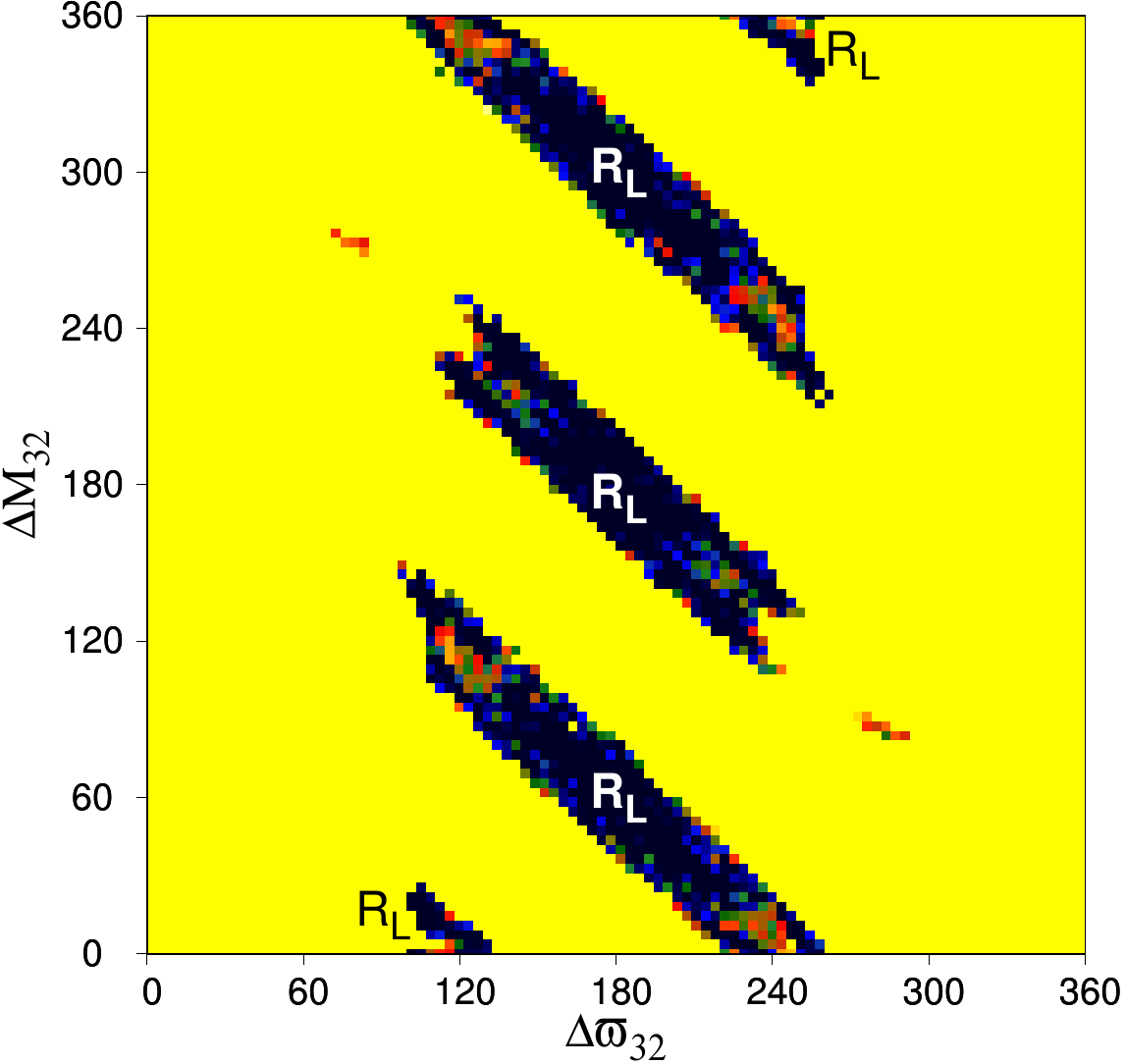}\vspace{-0.5cm}\\
\end{array}$
\begin{center}
\includegraphics[width=0.225\textwidth]{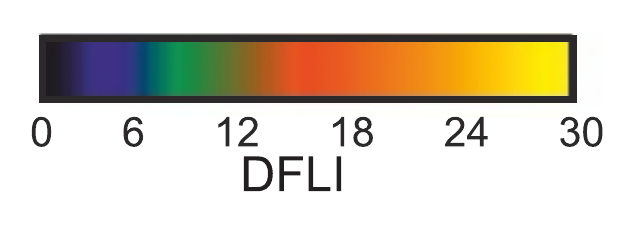}\vspace{-0.5cm}
\end{center}
\caption{DS-maps on the $(e_1,e_2)$ and $(e_2,e_3)$ planes, where the family $S_6$ is projected, and the $(\Delta\varpi_{21},M_{21})$ and $(\Delta\varpi_{32},M_{32})$ planes. Kepler-51 is identified by the magenta "$+$" symbol, while the errors are delineated by magenta dashed lines. The color bar illustrates the DFLI values; dark colors correspond to regular evolution, while pale ones signify chaoticity. $R_L$ indicates that both pairs of planets are locked in a two-body MMR (2/1 and 3/2 MMR), while $R_{S,L}$ indicates the 1/1 secondary resonance inside the 2/1 MMR for the inner pair and a locking in the 3/2 MMR for the outer pair of planets.}\label{S6_maps}
\end{figure*}

\begin{figure*}[tp!]%[ht!]
$\begin{array}{cc}
\includegraphics[width=0.48\textwidth]{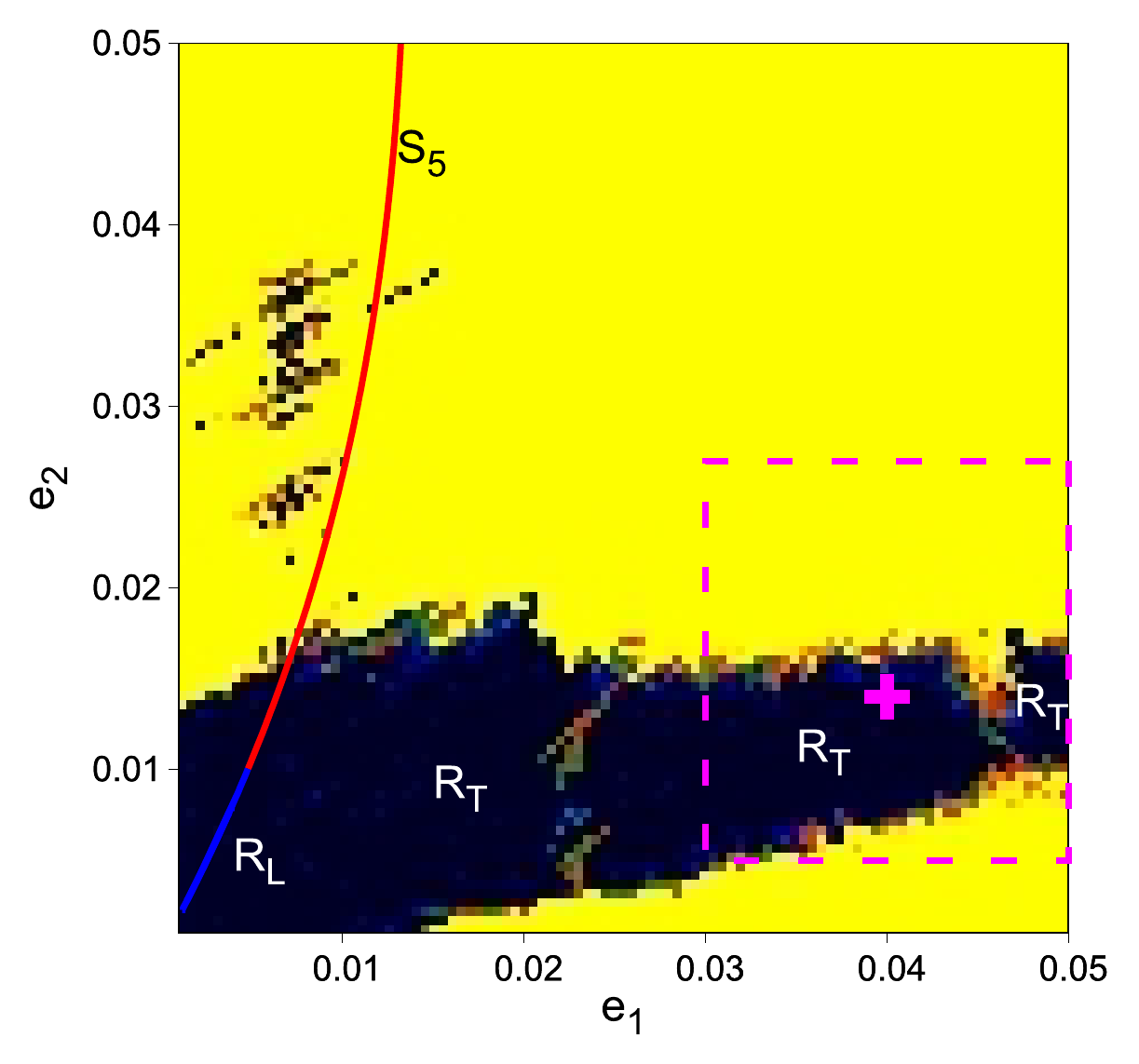}&\includegraphics[width=0.48\textwidth]{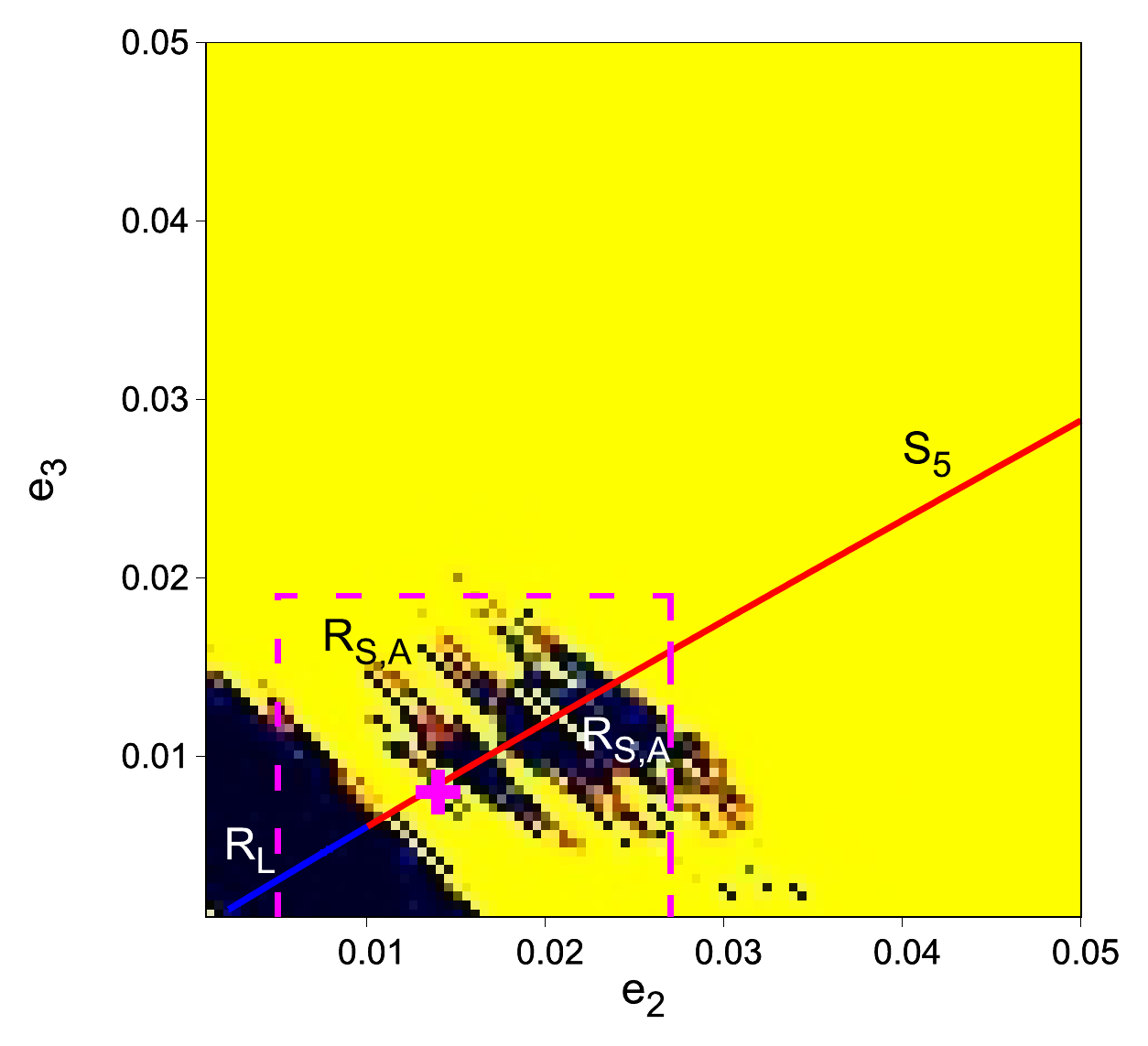}\vspace{-0.5cm}
\end{array}$
\begin{center}
\includegraphics[width=0.225\textwidth]{bar.eps}\vspace{-0.5cm}
\end{center}
\caption{DS-maps on the eccentricity planes $(e_1,e_2)$ and $(e_2,e_3),$ where the family $S_5$ is projected. $R_L$ denotes a two-body MMR (2/1 and 3/2 MMR for the inner and outer pairs of planets), $R_T$ indicates a three-body Laplace-like resonance (1:2:3 resonant chain), while $R_{S,A}$ indicates the 1/1 secondary resonance inside the 2/1 MMR for the inner pair and an apsidal difference oscillation/rotation observed for the outer pair of planets. Colors and lines as in Fig. \ref{S6_maps}.}\label{S5_maps}
\end{figure*}

\begin{figure*}[tp!]%[ht!]
$\begin{array}{cc}
\includegraphics[width=0.48\textwidth]{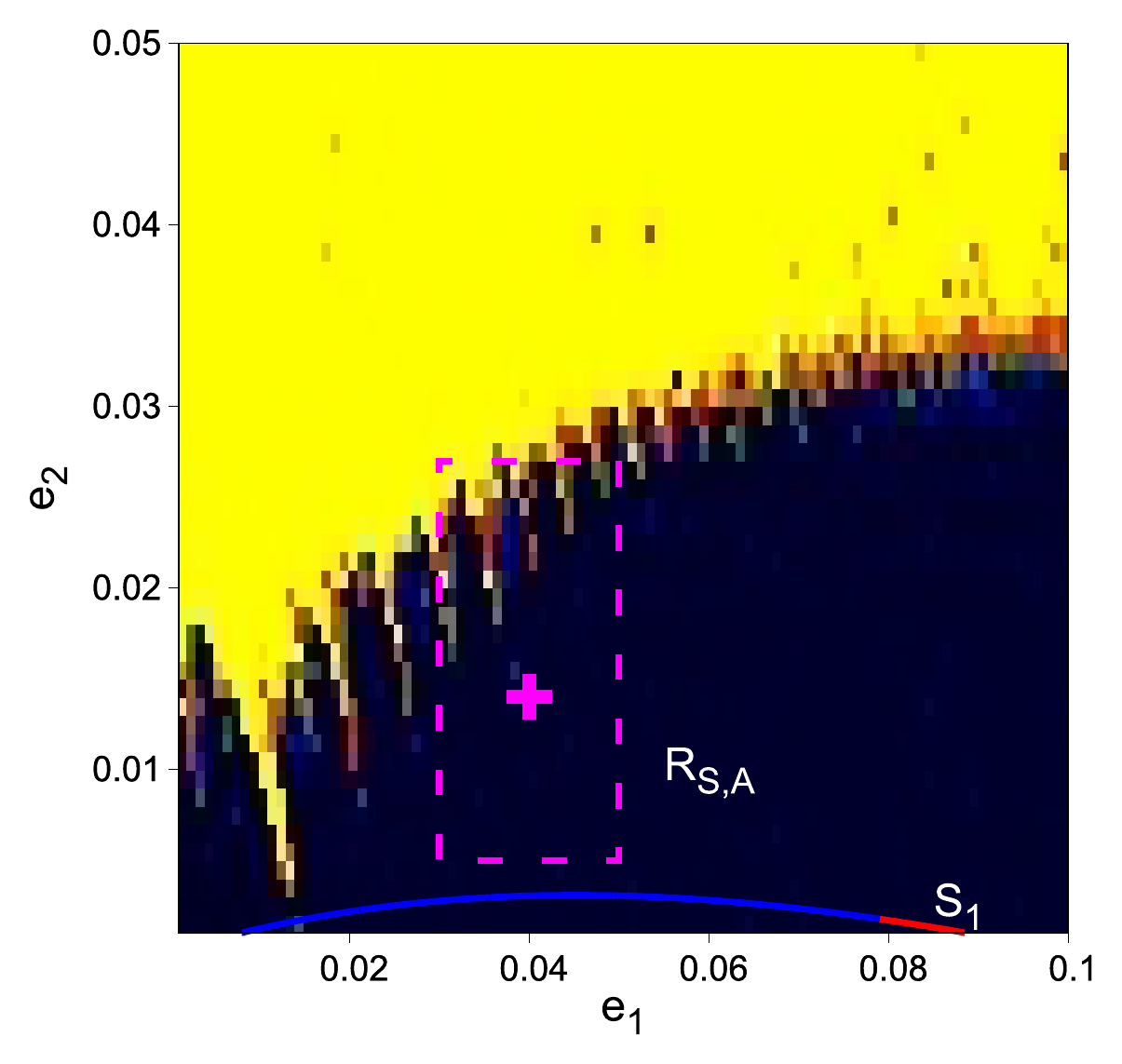}&\includegraphics[width=0.48\textwidth]{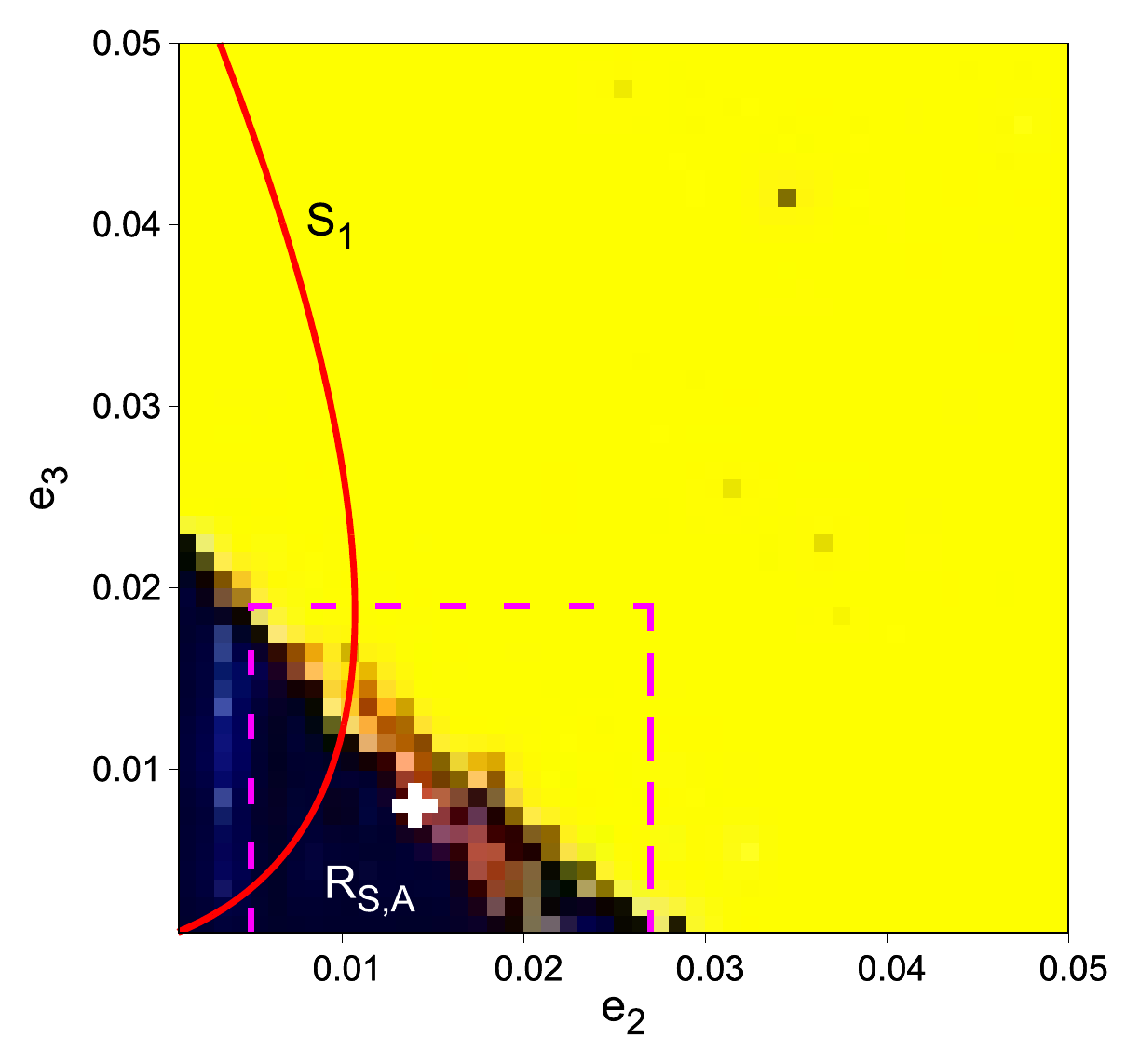}\vspace{-0.5cm}
\end{array}$
\begin{center}
\includegraphics[width=0.225\textwidth]{bar.eps}\vspace{-0.5cm}
\end{center}
\caption{DS-maps on the eccentricity planes $(e_1,e_2)$ and $(e_2,e_3)$, where the family $S_1$ is projected. $R_{S,A}$ indicates the 1/1 secondary resonance inside the 2/1 MMR for the inner pair and an apsidal difference oscillation/rotation observed for the outer pair of planets. Colors and lines as in Fig. \ref{S6_maps}.}\label{S1_maps}
\end{figure*}

\begin{figure}[ht!]
\includegraphics[width=0.5\textwidth]{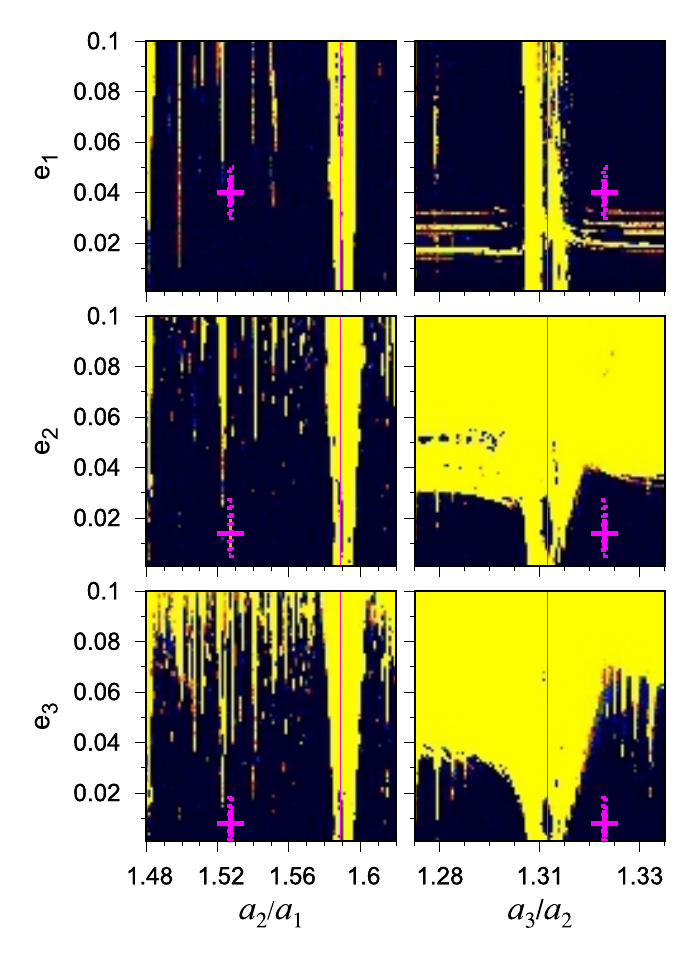}\vspace{-0.5cm}
\begin{center}
\hspace{1.25cm}\includegraphics[width=0.225\textwidth]{bar.eps}\vspace{-0.5cm}
\end{center}
\caption{DS-maps that showcase the extent of the 2/1 (left panels) and the 3/2 (right panels) MMRs in relation to each eccentricity value. The distinct $R_L$ regions appear at the values $\frac{a_2}{a_1}\approx 1.58$ and $\frac{a_3}{a_2}\approx 1.31$. The magenta lines showcase the nominal values of each MMR. The system Kepler-51 (magenta dashed lines enclosing the "$+$" symbol) is found at $\frac{a_c}{a_b}\approx 1.527$ and $\frac{a_d}{a_c}\approx 1.325$, where an apsidal resonance takes place. Colors and lines as in Fig. \ref{S6_maps}.}\label{f6_a2a3}
\end{figure}

\begin{figure}[ht!]
\includegraphics[width=0.5\textwidth]{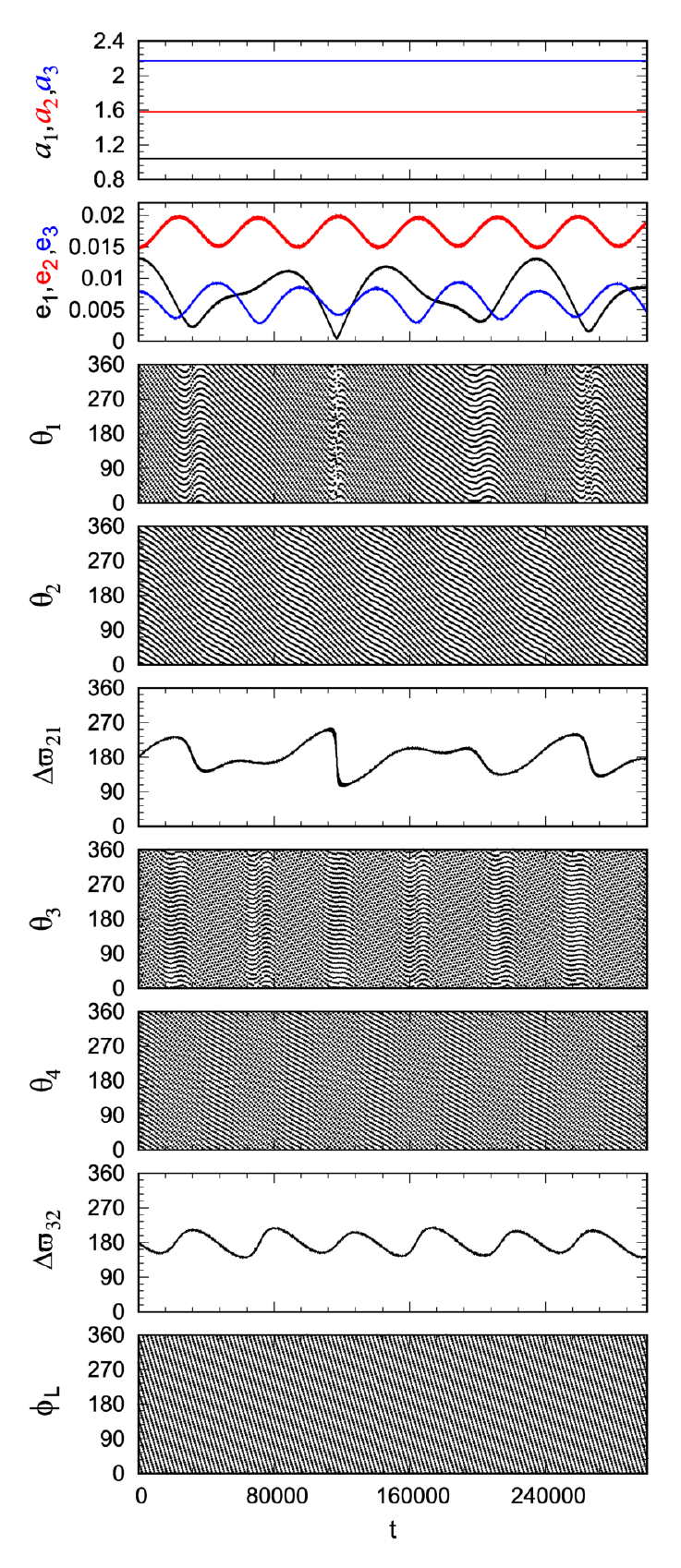}
\caption{Evolution of the orbital elements and resonant angles along an orbit initiated by the non-resonant low-eccentricity orbits, showcasing an apsidal resonance (shown in Fig. \ref{f6_a2a3}).}\label{f6_aaa}
\end{figure}

\begin{figure*}[h!]%[ht!]
$\begin{array}{cc}
\includegraphics[width=0.48\textwidth]{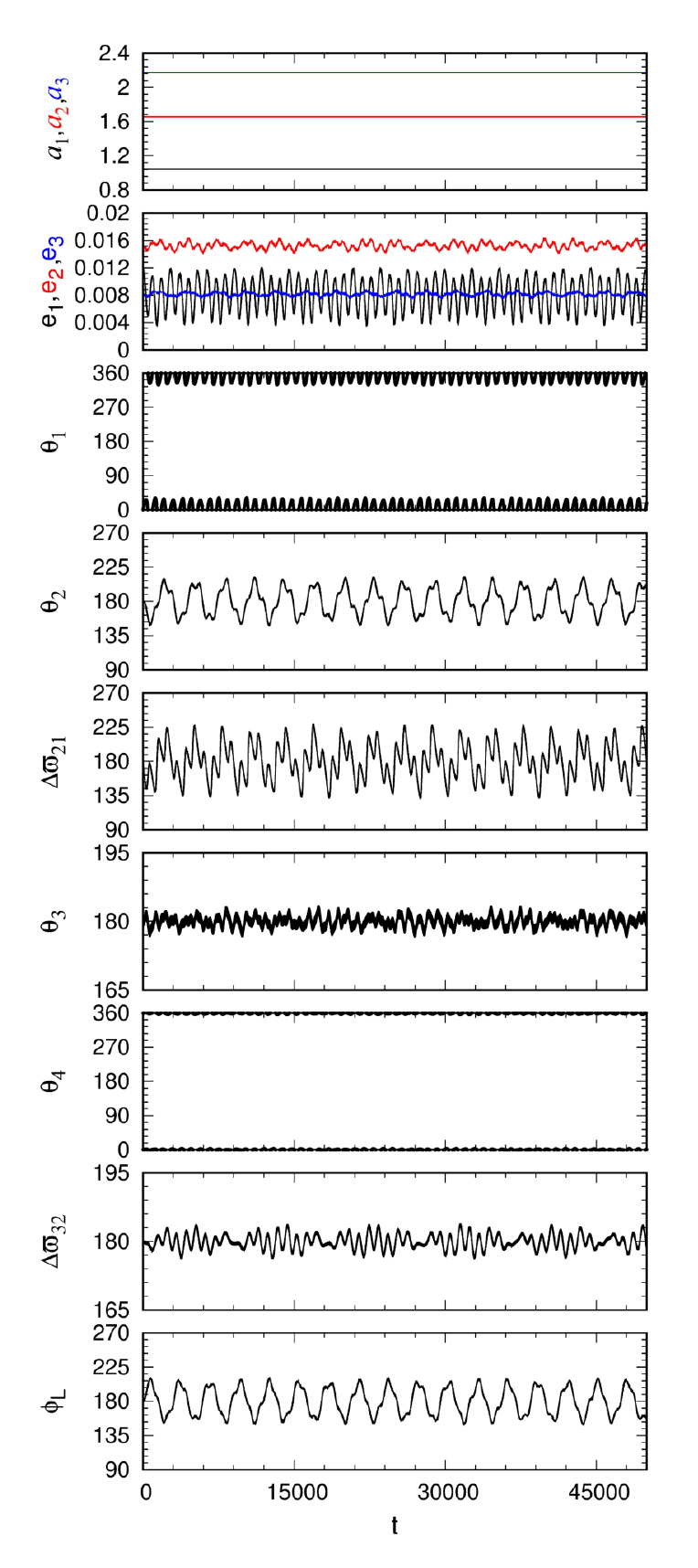}&\includegraphics[width=0.48\textwidth]{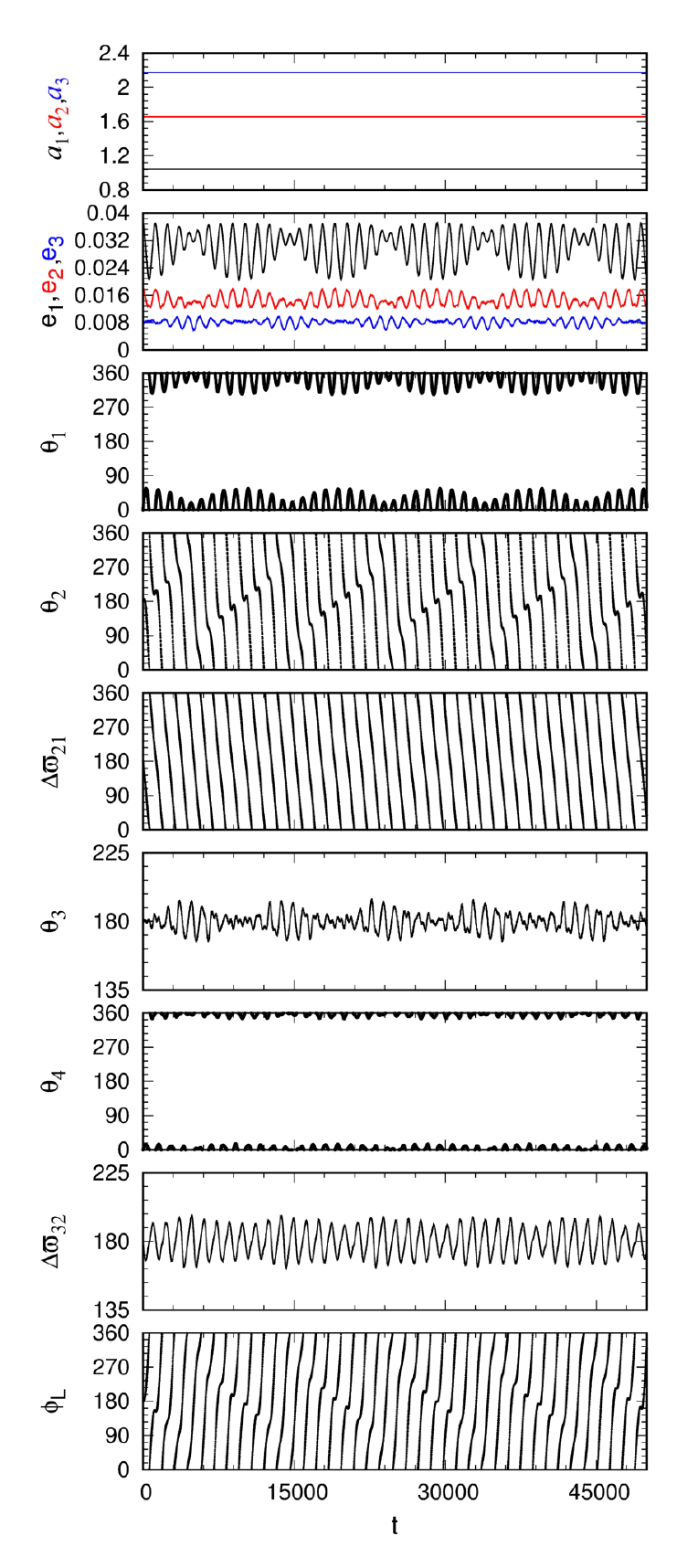}
\end{array}$
\caption{Evolution of the orbital elements and resonant angles along orbits initiated by the $R_L$ region (left panel) and the $R_{S,L}$ region (right panel) dynamically unveiled by the  $S_6$ family (shown in Fig. \ref{S6_maps}).}\label{S6_RL_RSL}
\end{figure*}

With regard to the computation of the DS-maps, we constructed $100\times 100$ grid planes and focused on the  $S_1$, $S_5$, and $S_6 $ families, namely, we chose specific orbital elements from the symmetric periodic orbits that belong to these families (their angles are reported in Table \ref{tconf}), since their stable segments are closer to the observational eccentricity values of Kepler-51. We remind the reader that the chosen families of periodic orbits have been computed for the masses of Kepler-51 (identified by label 1; i.e., \citet{masuda} in Fig. \ref{fams_all_masses} and shown in their full extent in Fig. \ref{123fams_z}) and for the 1:2:3 resonant chain. We note also that the values of the semimajor axes vary slightly along the families, but the resonance remains almost constant (see Appendix \ref{ApFams} for the differences between these elliptic $S_i$ families and the circular family along which the resonance varies). More precisely, on each grid plane, we vary a pair of the orbital elements, while keeping the rest of the orbital elements and masses of the periodic orbit fixed. In the following, we mention the orbital elements of the selected periodic orbits used for the construction of each DS-map and justify their selection.

In order to classify the orbits, we chose a maximum integration time for the computation of the DFLI equal to $t_{\rm max}=3Myr$, which corresponds approximately to 24 million orbits of the innermost planet, $b,$ or 8.5 million orbits of the outermost planet, $d,$ and was deemed appropriate, in order to distinguish chaos from order reliably for the particular application. We used the Bulirsch–Stoer integrator with a tolerance of $10^{-14}$. For small values of DFLI (dark-colored domains), a regular evolution for the planets is expected. In our study, we halt the integrations either when DFLI$(t)>30$ or when $t_{\rm max}$ is reached.

In Fig. \ref{S6_maps}, we present DS-maps on the planes $(e_1,e_2)$ and $(e_2,e_3)$ in the top panels and $(\Delta\varpi_{21},M_{21})$ and $(\Delta\varpi_{32},M_{32})$ in the bottom panels. The rest of the orbital elements that remain fixed for each of the initial conditions on the DS-maps are chosen from the planar periodic orbit in the $S_6$ family, with values for $e_2$ and $e_3$ that coincide approximately with the observational eccentricities, namely, $e_c$ and $e_d$ (depicted by the magenta "$+$" symbol in the top right panel), respectively. The selected periodic orbit has the following orbital elements: $a_1=1.04268089$, $a_2=1.65662643$, $a_3=2.17192389$, $e_1=0.0075847$, $e_2=0.0150072$, $e_3=0.0080335$, $\varpi_1=0^{\circ}$, $\varpi_2=180^{\circ}$, $\varpi_3=0^{\circ}$, $M_1=0^{\circ}$, $M_2=0^{\circ}$, and $M_3=180^{\circ}$, with the following configuration: $(\theta_1,\theta_2,\theta_3,\theta_4)$=$(0,\pi,\pi,0)$.

In Fig. \ref{S5_maps}, we present DS-maps by varying the eccentricities on the planes $(e_1,e_2)$ and $(e_2,e_3)$. The rest of the orbital elements, which remain fixed for each of the initial conditions on the DS-maps, are selected from a planar stable (blue) periodic orbit of the $S_5$ family being closer to the observational values $e_c$ and $e_d$ (errors counted in). Its orbital elements are: $a_1=0.99089944$, $a_2=1.56964834$, $a_3=2.05151434$, $e_1=0.0025747$, $e_2=0.0050208$, $e_3=0.0030763$, $\varpi_1=180^{\circ}$, $\varpi_2=180^{\circ}$, $\varpi_3=0^{\circ}$, $M_1=180^{\circ}$, $M_2=180^{\circ}$, and $M_3=0^{\circ}$, which correspond to the following configuration: $(\theta_1,\theta_2,\theta_3,\theta_4)$=$(\pi,\pi,0,\pi)$. 

Likewise, in Fig. \ref{S1_maps}, we present DS-maps by varying the eccentricities on the planes $(e_1,e_2)$ and $(e_2,e_3)$. The rest of the orbital elements, which remain fixed for each of the initial conditions on the DS-maps, are selected from a planar stable periodic orbit of the $S_1$family, where $e_1\approx e_b$. Its orbital elements are: $a_1=1.02641224$, $a_2=1.62959097$, $a_3=2.13398349$, $e_1=0.0400329$, $e_2=0.0030374$, $e_3=0.0000763$, $\varpi_1=0^{\circ}$, $\varpi_2=180^{\circ}$, $\varpi_3=0^{\circ}$, $M_1=0^{\circ}$, $M_2=180^{\circ}$, and $M_3=0^{\circ}$, with the following configuration: $(\theta_1,\theta_2,\theta_3,\theta_4)$=$(0,\pi,0,\pi)$. 

In Fig. \ref{f6_a2a3}, we provide the DS-maps that showcase the extent of each resonance (the 2/1 MMR in the left panels and the 3/2 MMR in the right ones) in relation to each planetary eccentricity value. We were guided by the same periodic orbit of the $S_6$ family that was used in Fig.~\ref{S6_maps} in the configuration $(\theta_1,\theta_2,\theta_3,\theta_4)$=$(0,\pi,\pi,0)$. The observational values of the semimajor axes and eccentricities are denoted by the magenta "$+$" symbol.

\begin{figure*}[h!]%[ht!]
$\begin{array}{cc}
\includegraphics[width=0.48\textwidth]{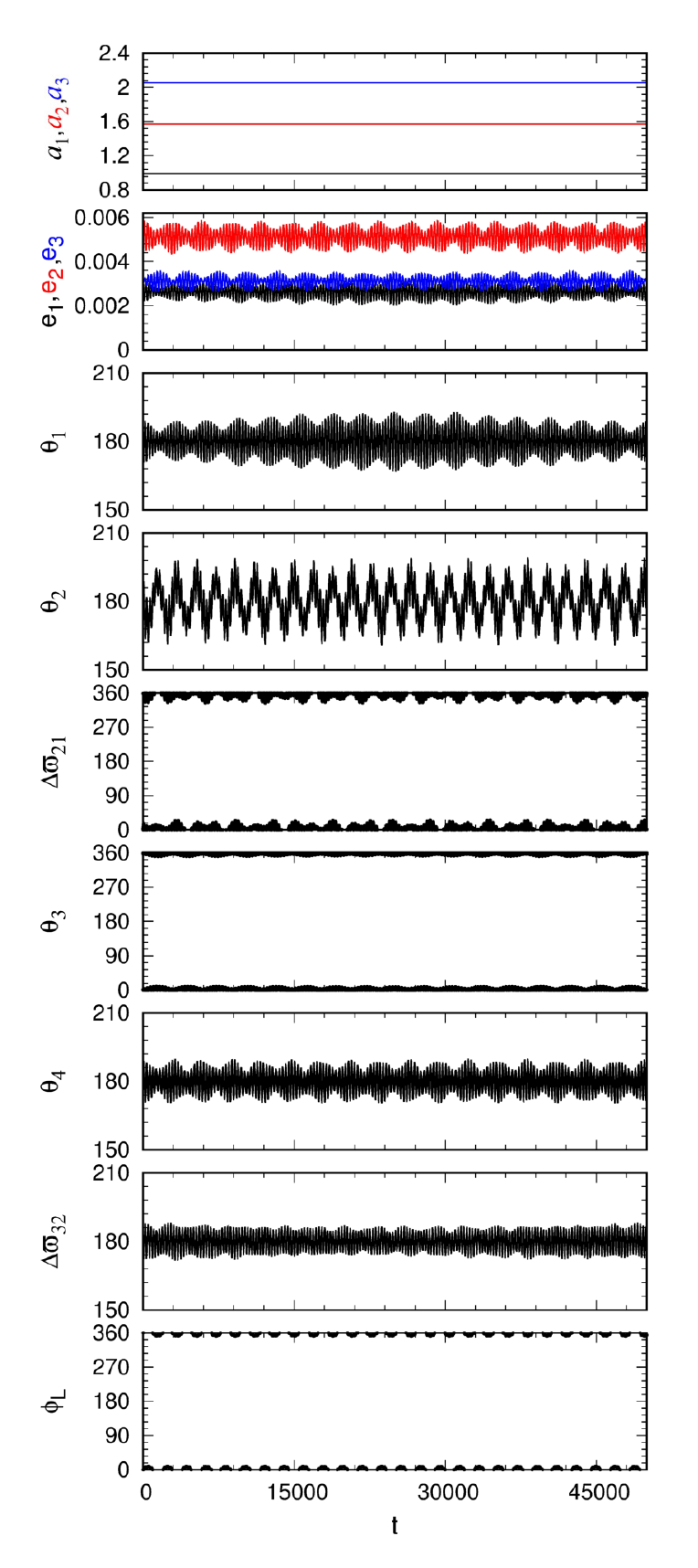}&\includegraphics[width=0.48\textwidth]{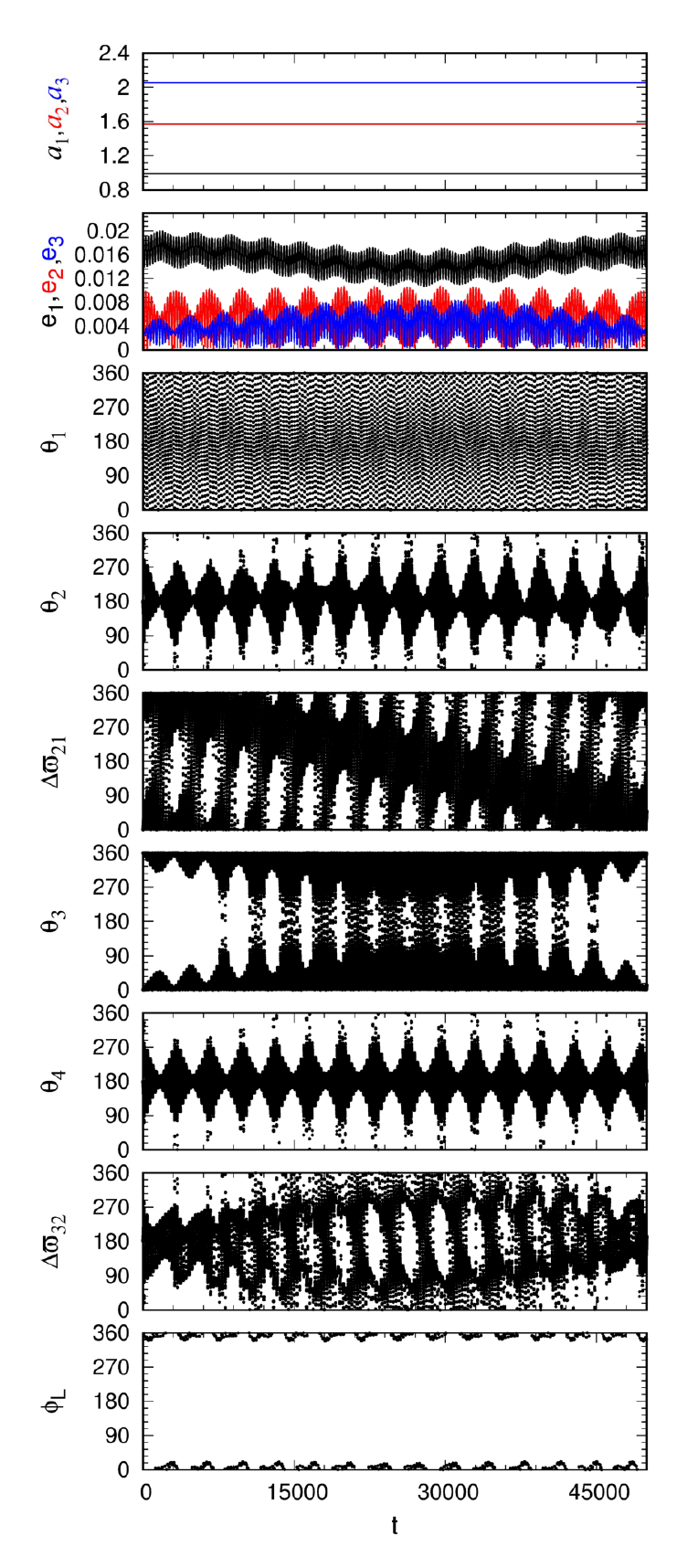}
\end{array}$
\caption{Evolution of the orbital elements and resonant angles along orbits initiated by the $R_L$ region (left panel) and the $R_T$ region (right panel) dynamically unveiled by the  $S_5$ family (shown in Fig. \ref{S5_maps}).}\label{S5_RL_RT}
\end{figure*}

\begin{figure*}[h!]%[ht!]
$\begin{array}{cc}
\includegraphics[width=0.48\textwidth]{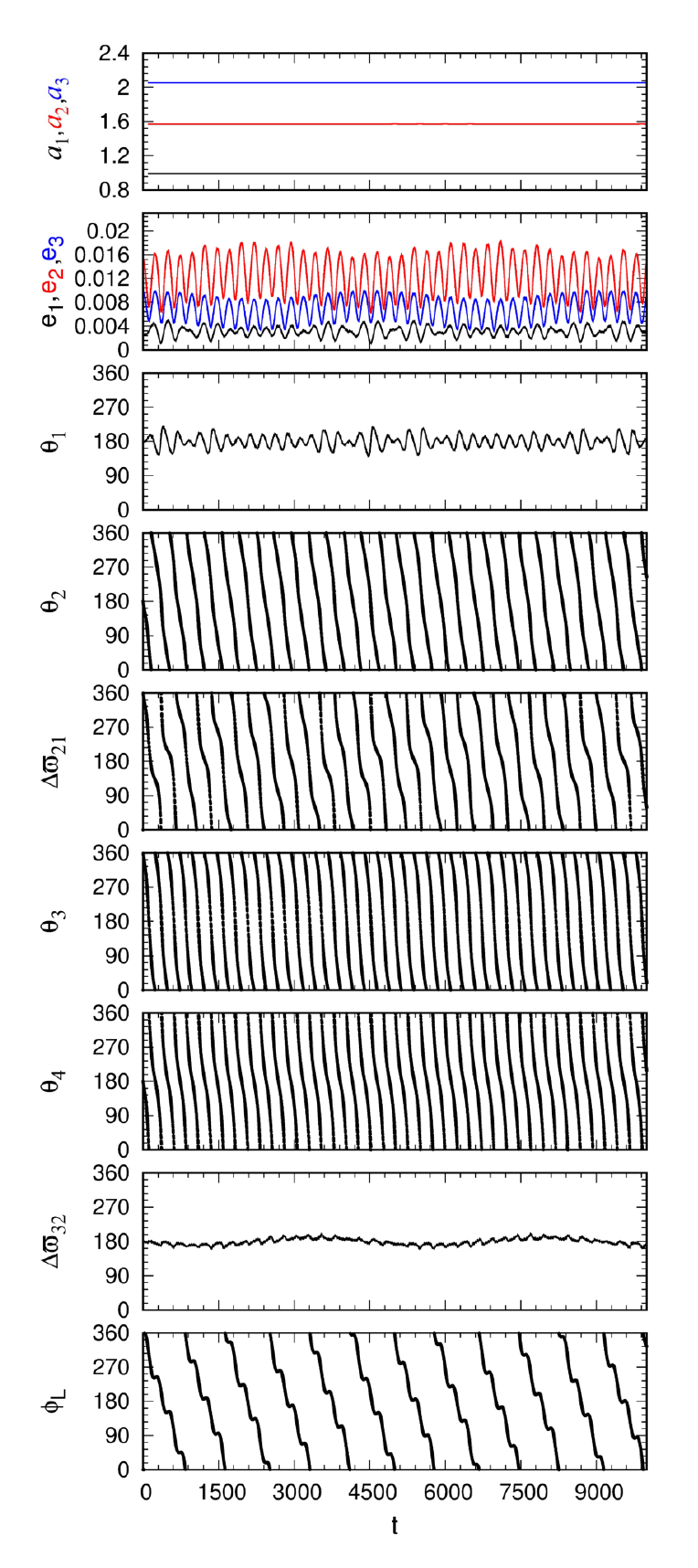}&\includegraphics[width=0.48\textwidth]{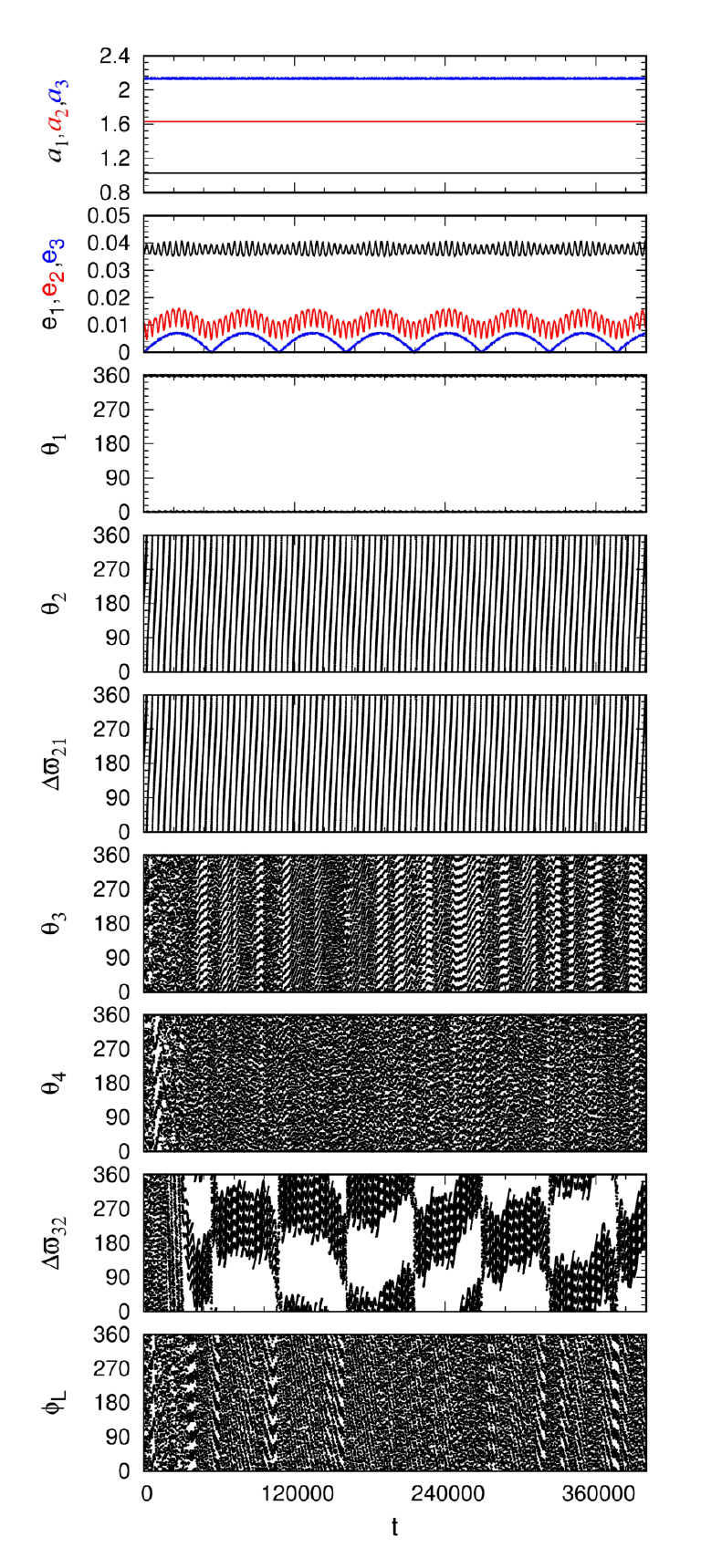}
\end{array}$
\caption{Evolution of the orbital elements and resonant angles along orbits initiated by  the $R_{S,A}$ regions dynamically unveiled by the $S_5$ (left panel) and $S_1$ (right panel) families (shown in Figs. \ref{S5_maps} and \ref{S1_maps}, respectively).}\label{S5_S1_RSA}
\end{figure*}

\subsubsection{Apsidal resonance for non-resonant low-eccentricity orbits}
At the low-eccentricity non-resonant orbits presented in Fig.~\ref{f6_a2a3}, that is, away from the distinct $R_L$ regions  found around the values $\frac{a_2}{a_1}\approx 1.58$ and $\frac{a_3}{a_2}\approx 1.31$, an apsidal resonance is found at the position of Kepler-51. In such an evolution (shown in Fig.~\ref{f6_aaa}), all the resonant angles rotate, but the apsidal differences librate about $\pi$, given the configuration $(\theta_1,\theta_2,\theta_3,\theta_4)$=$(0,\pi,\pi,0)$ of the periodic orbit from the  $S_6$ family used.

\subsubsection{Two-body MMRs, $R_L$}
 
The dynamical mechanism of two independent two-body MMRs, namely the 2/1 for the innermost planets and the 3/2 for the outermost ones, was encountered in the neighborhood of families $S_6$ and $S_5$ ($R_L$ symbols on the DS-maps in Figs. \ref{S6_maps} and \ref{S5_maps}). The evolutions of the orbital elements and the resonant angles from the initial conditions taken from these two "$R_L$ regions" are demonstrated in the left panels of Figs. \ref{S6_RL_RSL} and \ref{S5_RL_RT}. In the former case, the resonant angles, $(\theta_1,\theta_2,\theta_3,\theta_4)$, librate about $(0,\pi,\pi,0)$, while in the latter, the resonant angles librate about $(\pi,\pi,0,\pi)$.

Regarding the $S_5 $ family, this mechanism is stronger the closer we get to the family, as the libration widths of the resonant angles are very small. Therefore, we may conjecture that the two-body MMRs cannot act as a safeguard for the regular evolution of Kepler-51 when found in its dynamical vicinity.

In contrast, the  $S_6$  family may provide a region that could host Kepler-51 for long-time spans should the pairs be in two-body MMRs. More precisely, the dark-colored regular domain on the $(e_2,e_3)$ plane of Fig. \ref{S6_maps} almost engulfs the observational eccentricities, $e_c$ and $e_d$, along with their deviations. Given this overlap, a constraint should be imposed on the eccentricity $e_b$, since the $R_L$ region takes place when $e_1<0.02$, which is by far lower than the lowest deviation boundary (magenta dashed line) on the $(e_1,e_2)$ plane of Fig. \ref{S6_maps}. Additionally, guided by the same stable periodic orbit, the boundaries for the mean anomalies and apsidal differences can be deduced by the bottom panels of Fig. \ref{S6_maps}.

\subsubsection{Three-body Laplace-like resonance, $R_T$}

The dynamical mechanism of a three-body Laplace-like resonance, namely, the 1:2:3 resonant chain, was encountered in the neighborhood of the $S_5$ family ($R_T$ symbols in the left panel of Fig. \ref{S5_maps}). The evolution of the orbital elements and the resonant angles, with initial conditions taken from the "$R_T$ region" within the magenta dashed lines, is shown in the right panel of Fig. \ref{S5_RL_RT}, where the resonant angles rotate, but the Laplace angle librates about $0^{\circ}$. 

We may cautiously impose further constraints on the observational eccentricities for the 1:2:3 resonant chain to be long-term viable, by taking into account the regular domains (dark "$R_T$ regions") and the overlapping deviations (magenta dashed lines). More particularly, $e_c<0.016$ (deduced from the regular $R_T$ region in the left panel of Fig. \ref{S5_maps}), while $e_d<0.006$ (which is the highest value of $e_3$ in the stable segment of the  $S_5$ family).

\subsubsection{Combination of secondary resonance, two-body MMR and apsidal difference oscillation, $R_{S,L}$, and $R_{S,A}$}

The dynamical mechanism of the 1/1 secondary resonance inside the 2/1 MMR for the innermost planets and a locking in the 3/2 MMR for the outermost ones, was encountered in the neighborhood of the $S_6$ family ($R_{S,L}$ symbols in the top left panel of Fig. \ref{S6_maps}). The evolution of the orbital elements and the resonant angles, with initial conditions taken from the $R_{S,L}$ region within the magenta dashed lines, is shown in the right panel of Fig. \ref{S6_RL_RSL}, where a libration is observed for $\theta_1$ about $0^{\circ}$, $\theta_3$ about $180^{\circ}$, $\theta_4$ about $0^{\circ}$ and $\Delta\varpi_{32}$ about $180^{\circ}$ and a rotation for the rest of angles. 

The regular domain exhibiting this mechanism is particularly thin and the region delineated by the deviations of observational eccentricities (magenta dashed lines) is populated mainly by chaotic orbits. Hence, we do not put any constraints on Kepler-51 stemming from the $R_{S,L}$ region dynamically unveiled by the  $S_6$ family, since such an evolution may not be probable.

The dynamical mechanism of the 1/1 secondary resonance inside the 2/1 MMR for the innermost planets and an apsidal difference oscillation/rotation observed for the outer pair planets, was encountered in the neighborhood of the $S_5$ and $S_1$ families ($R_{S,A}$ symbols in Figs. \ref{S5_maps} and \ref{S1_maps}). The evolution of the orbital elements and the resonant angles, with initial conditions taken from the $R_{S,A}$ regions within the magenta dashed lines of Figs. \ref{S5_maps} and \ref{S1_maps}, is shown in the left and right panels of Fig. \ref{S5_S1_RSA}, respectively. In the $R_{S,A}$ region originating from the $S_5$ family, $\theta_1$ librates about $180^{\circ}$ and $\Delta\varpi_{32}$ oscillates about $180^{\circ}$, following the configuration of the chosen periodic orbit that was used for the DS-map construction. As for the $R_{S,A}$ region established by the periodic orbit of the $S_1$ family, $\theta_1$ librates about $0^{\circ}$, while $\Delta\varpi_{32}$ alternates between large amplitude oscillations about $0^{\circ}$ and circulations, revealing the chaotic nature of this motion.

Regarding the $R_{S,A}$ region in the DS-map of Fig. \ref{S5_maps}, we observe that $e_c$ and $e_d$ along with their deviations fall mainly on the unstable (red colored) periodic orbits. Therefore, the long-term stability may be possible but not very probable within these very small regular domains. 

Concerning the $R_{S,A}$ region in the DS-maps of Fig. \ref{S1_maps}, we observe that $e_b$ and $e_c$, along with their deviations, fall entirely on the regular domain. As a result, we performed a search for possible boundaries for the mean anomalies and apsidal differences guided by the same stable periodic orbit used for the DS-maps on the eccentricity planes. We found that essentially all values in the domain $[0,180]$ yield a regular $R_{S,A}$ mechanism. However, we recall that the above values hold for eccentricity, namely, $e_d\approx 0$ (see the $S_1$ family).

\section{Conclusions}\label{fin}

Motivated by the increasing number of exoplanets evolving in (or close) to MMRs and resonant chains, we analyzed the long-term orbital stability of the system Kepler-51. We presented novel results regarding the 1:2:3 resonant symmetric periodic orbits of the G4BP, which was used as a model to provide hints on the dynamics of the three planets.

For the planetary masses of Kepler-51, only four families were found to possess stable segments in the low-eccentricity dynamical neighborhood of the system. These families are in the symmetric configurations $(\theta_1,\theta_2,\theta_3,\theta_4)$=$(0,\pi,0,\pi)$ and $(0,\pi,\pi,\pi)$ ($S_1$ family); $(\pi,0,0,0)$ ($S_4$ family); $(\pi,\pi,0,\pi)$ ($S_5$ family); and $(0,\pi,\pi,0)$ ($S_6$ family).\footnote{Asymmetric configurations in resonant chains may also exist, but indications of such stability domains in three-planet systems were found in moderate eccentricity values when Jovian masses were assumed for the three planets \citep{voyEpjst}.} 

Guided by the stable symmetric periodic orbits of these families, we computed DS-maps, which basically unraveled the three main dynamical mechanisms which secure the long-term orbital stability of the system Kepler-51. More precisely, the 2/1 and 3/2 two-body MMRs (denoted as $R_L$), the 1:2:3 three-body Laplace-like resonance (denoted as $R_T$), and a combination of mechanisms for the inner and outer pairs separately, namely, the 1/1 secondary resonance inside the 2/1 MMR for the inner pair with either a 3/2 MMR or an apsidal difference oscillation for the outer pair (denoted as $R_{S,L}$ and $R_{S,A}$, respectively).

Based on the regular domains demonstrated in the DS-maps, we put possible constraints on the observational eccentricities, mean anomalies, and apsidal differences. For the first scenario in the $R_L$ region, we concluded that $e_b<0.02$ and we presented possible islands of stability for the angles, so that such two-body MMRs endure for long time spans. For the second scenario in the $R_T$ region, we found  $e_c<0.016$ and $e_d<0.006$ to be viable for such a chain. For the third scenario, we deduced that an evolution comprising the 1/1 secondary resonance with a 3/2 MMR ($R_{S,L}$ region) may be possible but not probable. Additionally, we showed that an evolution in an $R_{S,A}$ region, despite it being of a chaotic nature, fits very well with the observational eccentricities $e_b$ and $e_c$ (and almost any value for the mean anomalies and the apsidal differences), as long as $e_d\approx 0$.

With regard to two-planet systems, many fitting methods have been developed and dynamical analyses have been performed for giant planets locked in MMRs in tandem with migration simulations \citep[see e.g.,][]{Hadden2020}. An efficient fitting of the observational data for systems of three planets in (or near) a resonance is beyond any question.  With the aim of obtaining an optimum deduction of the orbital elements, this study exemplifies the need for dynamical analyses based on periodic orbits  performed in parallel to the data fitting methods.

\begin{acknowledgements} We are grateful to an anonymous reviewer whose thorough remarks and suggestions immensely improved our study. The research of KIA is co-financed by Greece and the European Union (European Social Fund - ESF) through the Operational Programme ``Human Resources Development, Education and Lifelong Learning'' in the context of the project ``Reinforcement of Postdoctoral Researchers - $2^{\rm{nd}}$ Cycle'' (MIS-5033021), implemented by the State Scholarships Foundation (IKY). Results presented in this work have been produced using the Aristotle University of Thessaloniki (AUTh) High Performance Computing Infrastructure and Resources. \end{acknowledgements}

\bibliographystyle{aa}
\bibliography{123_kep51}

%\clearpage

\begin{appendix}

\section{Hadjidemetriou's approach to the planar 4-body problems}\label{model}

\subsection{Equations of Motion}\label{eqsm}

Let us first consider a system of 4 bodies, $p_i$ $(i=0,..,3)$, with masses $m_i$ $(i=0,..,3)$, respectively, which move under their mutual gravitational attraction in the inertial frame $XGY$, where $G$ is their center of mass and $\overline{R}_i$ the position vector with respect to $G$. Then, we define a rotating frame of reference, $xOy$, whose origin, $O$, is the center of mass of $p_0$ and $p_1$, which move on the $Ox$-axis (positive direction from $p_0$ to $p_1$). The position vector for each body in this frame with respect to $O$ is $\overline{r}_i$, while $x_1, x_2, x_3, y_2, y_3$ are the relative coordinates, $\dot x_1, \dot x_2, \dot x_3, \dot y_2, \dot y_3$ are the relative velocities, $\theta$ is the angle between the $Ox$ and $GX$ axes, and $\dot \theta$ the angular velocity. The motion takes place on a plane, while the planes $XGY$ and $xOy$ coincide. In this work, subscript 0 refers to the Star, while the total mass of the system, $m=\sum\limits_{i=0}^3 m_i$, and the gravitational constant, $G$, are normalized to unity, so that $m_0=1-m_1-m_2-m_3$ and $G=1$, respectively. 

Given the above we have
\begin{equation}
\overline{r}=\frac{1}{m_0+m_1} \sum\limits_{i=0}^3 m_i \overline{R}_i
,\end{equation}
with $\overline{r}=\overline{OG}$ 
and
\begin{equation}
x_0=-q x_1 
,\end{equation}
where $q=m_1/m_0$. Equivalently, we can define the position of $p_0$ and $p_1$ as 
\begin{eqnarray}
\begin{array}{l}
x_0=-(1-\mu_{01}) r \\
x_1= \mu_{01} r
\end{array}
\end{eqnarray}
where $\mu_{01}=\frac{m_0}{m_0+m_1}$ and $r=x_1-x_0$.

The Lagrangian of such a system in the General 4-Body Problem (G4BP) is
\begin{equation}\label{lagrange}
\begin{aligned}
\textsl{L}= & \frac{1}{2} m_1 \frac{m_0}{m_1 + m_0} \dot r^2 + 
 \frac{1}{2} m_2 (1 - m_2) (\dot x_2^2 + \dot y_2^2) + \\
& \frac{1}{2} m_3 (1 - m_3)(\dot x_3^2 + \dot y_3^2) - 
 m_2 m_3(\dot x_2 \dot x_3 + \dot y_2 \dot y_3) + \\
& \frac{1}{2} \dot \theta^2 [m_1 \frac{m_0}{m_1 + m_0} r^2 + m_2 (1 - m_2)(x_2^2 + y_2^2) + \\
& m_3 (1 - m_3)(x_3^2 + y_3^2) - 2 m_2 m_3 (x_2 x_3 + y_2 y_3)] +\\
& \dot \theta [m_2 (1 - m_2)(\dot y_2 x_2 - \dot x_2 y_2) +\\
& m_3 (1 - m_3)(\dot y_3 x_3 - \dot x_3 y_3) + \\
& m_2 m_3(\dot x_3 y_2 + \dot x_2 y_3 - \dot y_3 x_2 - \dot y_2 x_3)] - V
\end{aligned}
\end{equation}
where $V=\underset{i\neq j}{\sum\sum}\frac{m_i m_j}{r_{ij}}$ with $r_{ij}=\sqrt{(x_i-x_j)^2+(y_i-y_j)^2}$ $(i,j=0,...3)$.

It is evident from Eq.~\ref{lagrange} that $\theta$ is an ignorable variable $(\frac{\partial \textsl{L}}{\partial \theta}=0)$ and, hence, the angular momentum integral, $P_{th}=\frac{\partial \textsl{L}}{\partial \dot \theta}$, exists as such:
\begin{equation}\label{pth}
\begin{aligned}
P_{th}= & \dot \theta [ m_1 \mu_{01} \dot r^2 + m_2 (1 - m_2) (\dot x_2^2 + \dot y_2^2) + \\
& m_3 (1 - m_3)(\dot x_3^2 + \dot y_3^2) - \\
& 2 m_2 m_3( x_2  x_3 +  y_2 y_3) ] + \\
& m_2 (1 - m_2)(\dot y_2 x_2 - \dot x_2 y_2) + \\
& m_3 (1 - m_3)(\dot y_3 x_3 - \dot x_3 y_3) + \\
& m_2 m_3(\dot x_3 y_2 + \dot x_2 y_3 - \dot y_3 x_2 - \dot y_2 x_3).
\end{aligned}
\end{equation}

Therefore, we have a system of 5 degrees of freedom in the rotating frame with the following equations of motion:

\begin{subequations}\label{eqn:all-lines}
\begin{align}
&\ddot r =  r \dot \theta^2 + A \label{eqn:line-1}      ,\\
&\ddot x_2 =  \ddot \theta y_2 + 2 \dot \theta \dot y_2 + \dot \theta^2 x_2 + B_2 + P_2\label{eqn:line-2} ,\\
&\ddot y_2 = -\ddot \theta x_2 - 2 \dot \theta \dot x_2 + \dot \theta^2 y_2 + C_2 + Q_2\label{eqn:line-3} ,\\
&\ddot x_3 =  \ddot \theta y_3 + 2 \dot \theta \dot y_3 + \dot \theta^2 x_3 + B_3 + P_3\label{eqn:line-4} ,\\
&\ddot y_3 = -\ddot \theta x_3 - 2 \dot \theta \dot x_3 + \dot \theta^2 y_3 + C_3 + Q_3,\label{eqn:line-5}  
\end{align}
\end{subequations}
where
\begin{equation}\label{eqs_1}
\begin{aligned}
A= & -\frac{m_1+m_0}{r^2} + \\
& m_2 \frac{x_2 - \mu_{01} r}{r_{12}^3} - m_2 \frac{x_2 + (1-\mu_{01}) r}{r_{02}^3} + \\
& m_3 \frac{x_3 - \mu_{01} r}{ r_{13}^3} - m_3 \frac{x_3 + (1-\mu_{01}) r}{r_{03}^3} \\
B_2= & -(1-\mu_{01}) \frac{x_2 - \mu_{01} r}{r_{12}^3}   - \mu_{01} \frac{x_2 + (1-\mu_{01}) r}{r_{02}^3}  \\
B_3= & -(1-\mu_{01}) \frac{x_3 - \mu_{01} r}{r_{13}^3}   - \mu_{01} \frac{x_3 + (1-\mu_{01}) r}{r_{03}^3} \\
C_2= & -(1-\mu_{01}) \frac{y_2}{r_{12}^3} - \mu_{01} \frac{y_2}{r_{02}^3}\\
C_3= & -(1-\mu_{01}) \frac{y_3}{r_{13}^3} - \mu_{01} \frac{y_3}{r_{03}^3}\\
P_2= & m_3 (B_3 - B_2 + \frac{x_3 - x_2}{r_{32}^3})\\
P_3= & m_2 (B_2 - B_3 + \frac{x_2 - x_3}{r_{23}^3})\\
Q_2= & m_3 (C_3 - C_2 + \frac{y_3 - y_2}{r_{32}^3})\\
Q_3= & m_2 (C_2 - C_3 + \frac{y_2 - y_3}{r_{23}^3}),
\end{aligned}
\end{equation}
while the quantity $\ddot \theta$ is found by differentiating Eq.~\ref{pth} with respect to time.

For our numerical computations in the G4BP, we set $\dot \theta(0)=1$ and arbitrarily chose $\theta(0)=0$ without any loss of generality. 

In the restricted 4BP (R4BP), we consider the motion of the massless planet of $p_3$ ($m_3=0$), which does not affect the motion of the other three main bodies \citep[see, e.g.,][]{hadj4Body}. By taking the limit $m_3\rightarrow 0$ in the equations of motion (Eq.~\ref{eqn:all-lines}), Eqs.~\ref{eqn:line-4}, and \ref{eqn:line-5} uncouple from Eqs.~\ref{eqn:line-1}-\ref{eqn:line-3}, which now constitute the equations of motion of the three-body problem, and so we obtain the equations of motion in the R4BP.

\subsection{Symmetric periodicity conditions and continuation methods}\label{pc}

An orbit $\mathbf{X}(t)=(x_1(t),x_2(t),x_3(t),y_2(t),y_3(t),\dot x_1(t),\dot x_2(t),\dot x_3(t),\allowbreak \dot y_2(t),\dot y_3(t))$ is periodic of period $T$ in the G4BP if it satisfies the periodic conditions:
\begin{equation} \label{pocon}
\begin{array}{ll}
\dot{x}_1(T)=\dot{x}_1(0)=0, & \\
x_1(T)=x_1(0), & \\ 
x_2(T)=x_2(0) ,& y_2(T)=y_2(0), \\
x_3(T)=x_3(0) ,& y_3(T)=y_3(0), \\
\dot{x}_2(T)=\dot{x}_2(0), & \dot{y}_2(T)=\dot{y}_2(0),\\
\dot{x}_3(T)=\dot{x}_3(0), & \dot{y}_3(T)=\dot{y}_3(0). 
\end{array}
\end{equation}

A periodic orbit is symmetric with respect to the $Ox$-axis of the rotating frame if it remains invariant under the fundamental symmetry $\Sigma$ \citep[see e.g.,][]{hen97}: 
\begin{equation}\label{sigma}
\Sigma: (t,x,y)\rightarrow (-t,x,-y),
\end{equation}
which the system in Eq. \ref{lagrange} follows. Following \citet{hadjmich81}, we consider the initial conditions on a Poincar\'e surface of section $y_2=0$.  A symmetric periodic orbit crosses perpendicularly the section twice in one period and thus we obtain the following conditions, which are sufficient for the periodicity of the orbit with a period of $T=2t^*$:

\begin{equation}\label{a10}
\begin{array}{l}
y_3(0)=y_3(t^*)=0, \;  \dot x_1(0)= \dot x_1(t^*)=0, \\
\dot x_2(0)= \dot x_2(t^*)=0, \;  \dot x_3(0)= \dot x_3(t^*)=0,
\end{array}
\end{equation}
where $t^*$ is the time of the first section cross. The rest non-zero initial conditions form the five-dimensional space

\begin{equation}\label{spo}
\Pi_5=\{(x_1(0),x_2(0),x_3(0),\dot{y}_2(0),\dot{y}_3(0))\}
,\end{equation}
with $\dot{x}_1(0)=\dot{x}_2(0)=0=\dot{x}_3(0)=y_2(0)=y_3(0)=0$. 

In order to obtain a point in $\Pi_5$ that corresponds to a periodic orbit, we keep $x_1(0)$ fixed and we  computationally determine the rest four conditions, so that they satisfy Eq. \ref{a10}. Then, by varying $x_1(0)$ we compute a set of periodic orbits (a family) that forms a characteristic curve in $\Pi_5$.  These characteristic curves are presented as projections in planes of initial conditions of the rotating frame or, after conversion, in planes of orbital elements. 

The above continuation method can take place also by starting with zero planetary masses and then increasing them. However, it has been proven that such a continuation does not hold when the period of the periodic orbit, $T$, is a multiple of $2\pi$ or the mean-motion resonance is of the first order. In particular, following Eq.~\ref{PQ}, the continuation is not applied when $T=2\pi k$, namely, at the mean-motion resonances:

\begin{equation}\label{PQchain}
\frac{T_1}{T_2}=\frac{k-P}{k}, \;\; \frac{T_1}{T_3}=\frac{k-Q}{k}, \;\; k\in \mathbb{Z}^*.
\end{equation} 

The proof of the continuation methods for the $N$-body problems, along with the above-mentioned exceptions, can be found in \citet{hadj76,hadjNbody}.

\begin{figure}[h!]%[ht!]
\begin{center}
\includegraphics[width=0.5\textwidth]{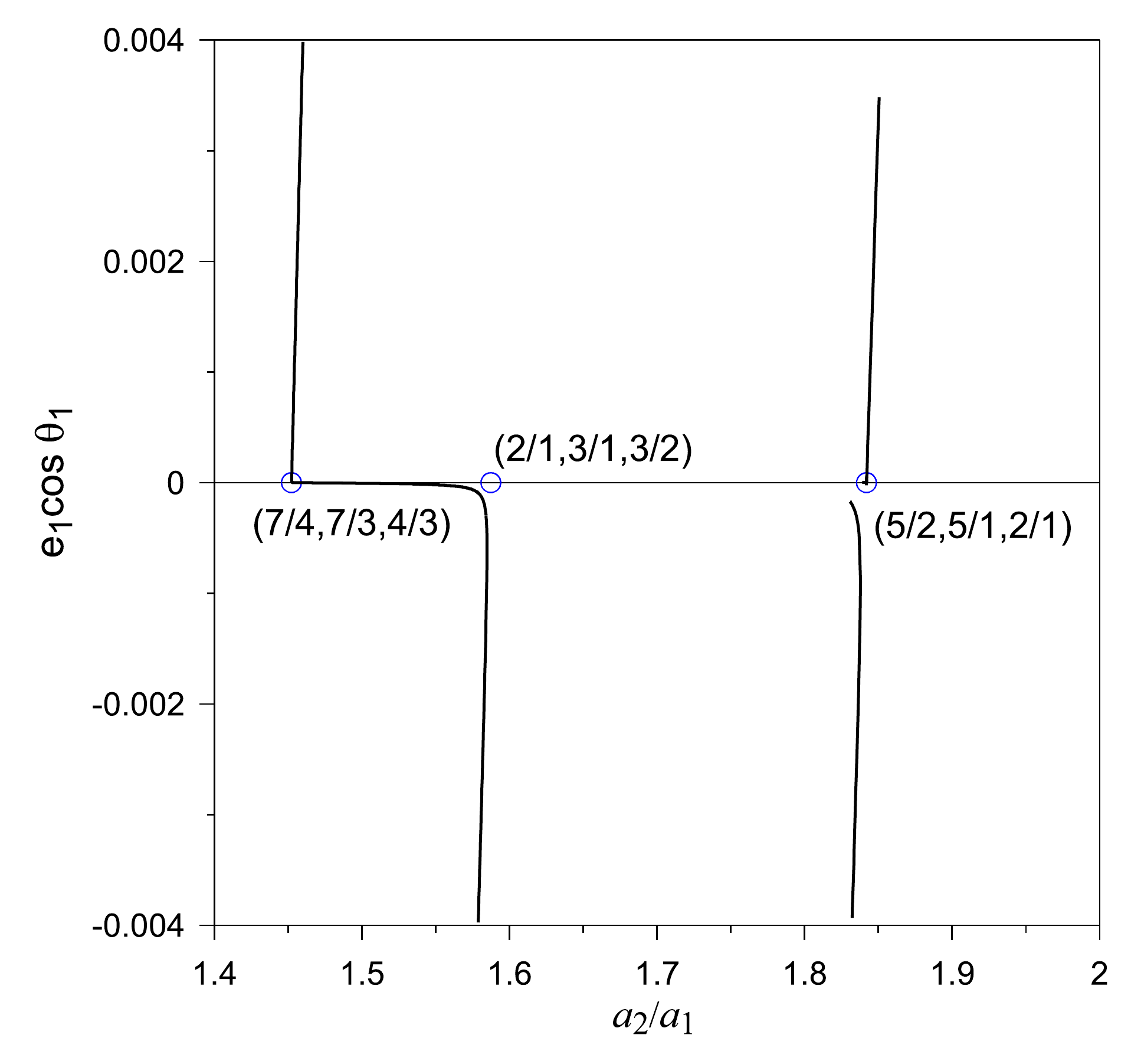}%\vspace{-0.4cm}
\end{center}
\caption{Circular family for the unperturbed case at $e_1=0$ and examples of families in the G4BP deflected from the resonant periodic orbits found by Eq.~\ref{PQchain} at $(\frac{T_2}{T_1},\frac{T_3}{T_1},\frac{T_3}{T_2})= (\frac{5}{2},\frac{5}{1},\frac{2}{1})\;{\rm for}\; k=3,\; (\frac{2}{1},\frac{3}{1},\frac{3}{2})\;{\rm for}\; k=4,\;{\rm and}\; (\frac{7}{4},\frac{7}{3},\frac{4}{3}) \;{\rm for}\; k=5$ (blue circles) computed for $m_1=m_2=m_3=10^{-6}$.}\label{pert}%\vspace{-0.48cm}
\end{figure}

\subsection{Continuation of periodic orbits in the 1:2:3 resonant chain}\label{ApFams}

In our study, we started from the degenerate (unperturbed) case, where all bodies move on circular Keplerian orbits and all masses are equal to zero. In this case, the resonant periodic orbits in the 1:2:3 resonant chain have a period $T=12\pi$ (see Eq.~\ref{PQ}). 

In Fig. \ref{pert}, we show the gaps that are formed at the first-order resonances when $m_i\neq 0$ $(i=1,2,3)$, between the bifurcation points (blue circles) of the circular family, along which the resonance varies, in the unperturbed case (found at $e_1=0$) and some of the families in the G4BP (found at $e_1\cos\theta_1\neq 0$). In other words, when we switched on the masses and performed a continuation with respect to the mass until the observational mass value for each planet of Kepler-51 was reached, the periodic orbits in the G4BP were not connected smoothly with the ones in the unperturbed case, since the continuation with respect to the mass cannot be applied (see Eq.~\ref{PQchain}). Then, we followed the continuation method to alter the $x_2$ variable of the system by keeping these mass values fixed. The families in the G4BP are deflected from the resonant periodic orbit in the unperturbed case, while the resonance along them remains almost constant. In this way, we obtained the families of symmetric periodic orbits in the 1:2:3 resonant chain up to high eccentricity values illustrated in Fig.~\ref{123fams_z}. In Poincar\'e's terminology, these are the periodic orbits of a "second kind."\footnote{The periodic orbits of the "first kind" are the ones of planetary type (non-zero masses) describing nearly circular motion.} The termination of the families takes place either at close encounters between at least one pair of planets or when the continuation method stalls, as the convergence to the periodicity conditions becomes very slow at very high eccentricity values.

We note that the bifurcation points of the circular family can generate different families of the same resonant chain, which differ in the phases of planetary configurations. These configurations correspond to the initial location at $t=0$ of the 3 bodies, namely $p_1$, $p_2,$ and $p_3$, on the $x$-axis and are reflected on the values of the longitudes of pericenter and the mean anomalies of the periodic orbits in the G4BP.

\begin{figure}
$\begin{array}{c}
\includegraphics[width=0.472\textwidth]{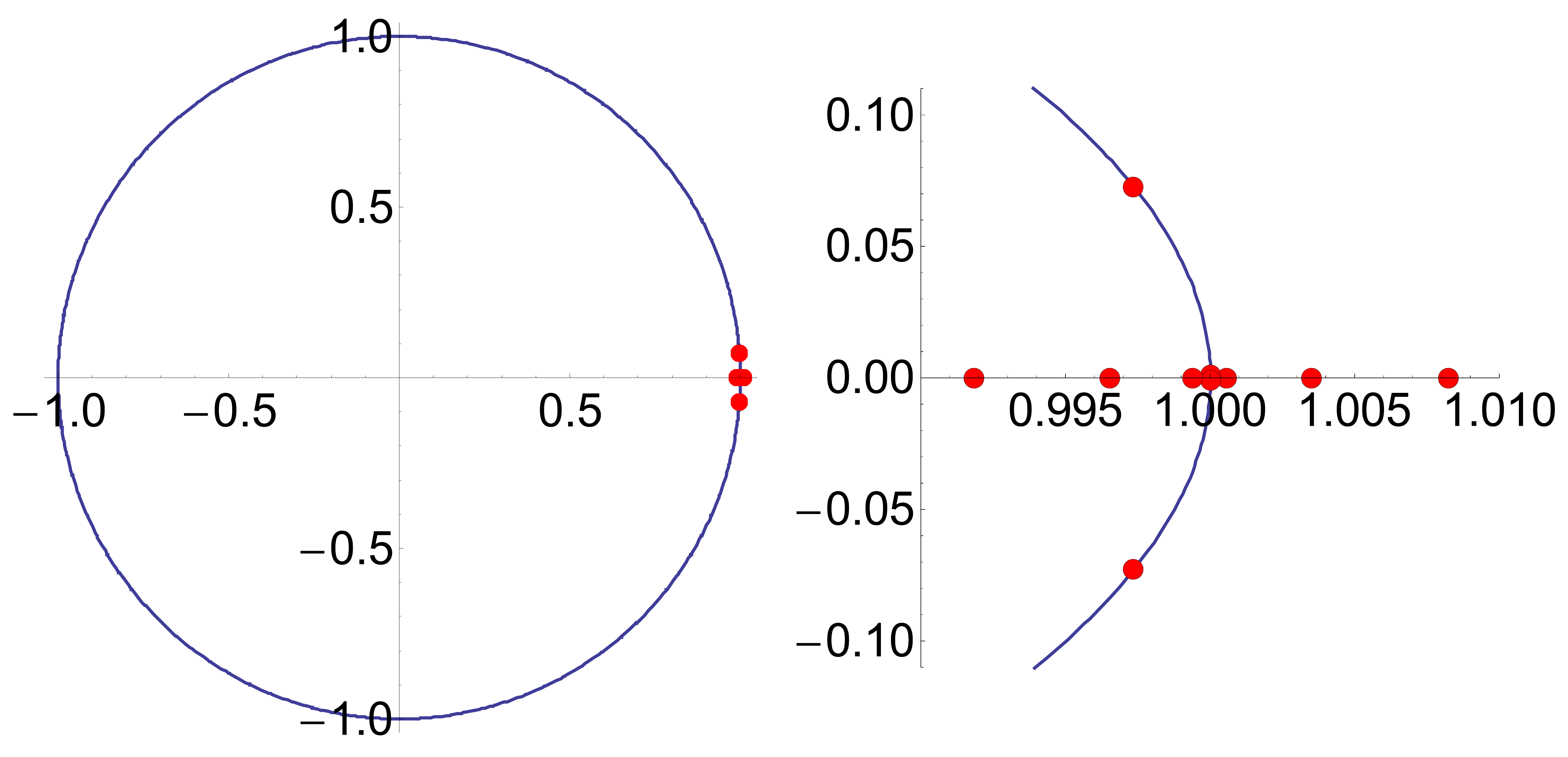}\hspace{-0.5cm}{\rm (a)}\vspace{-0.15cm}\\
\includegraphics[width=0.472\textwidth]{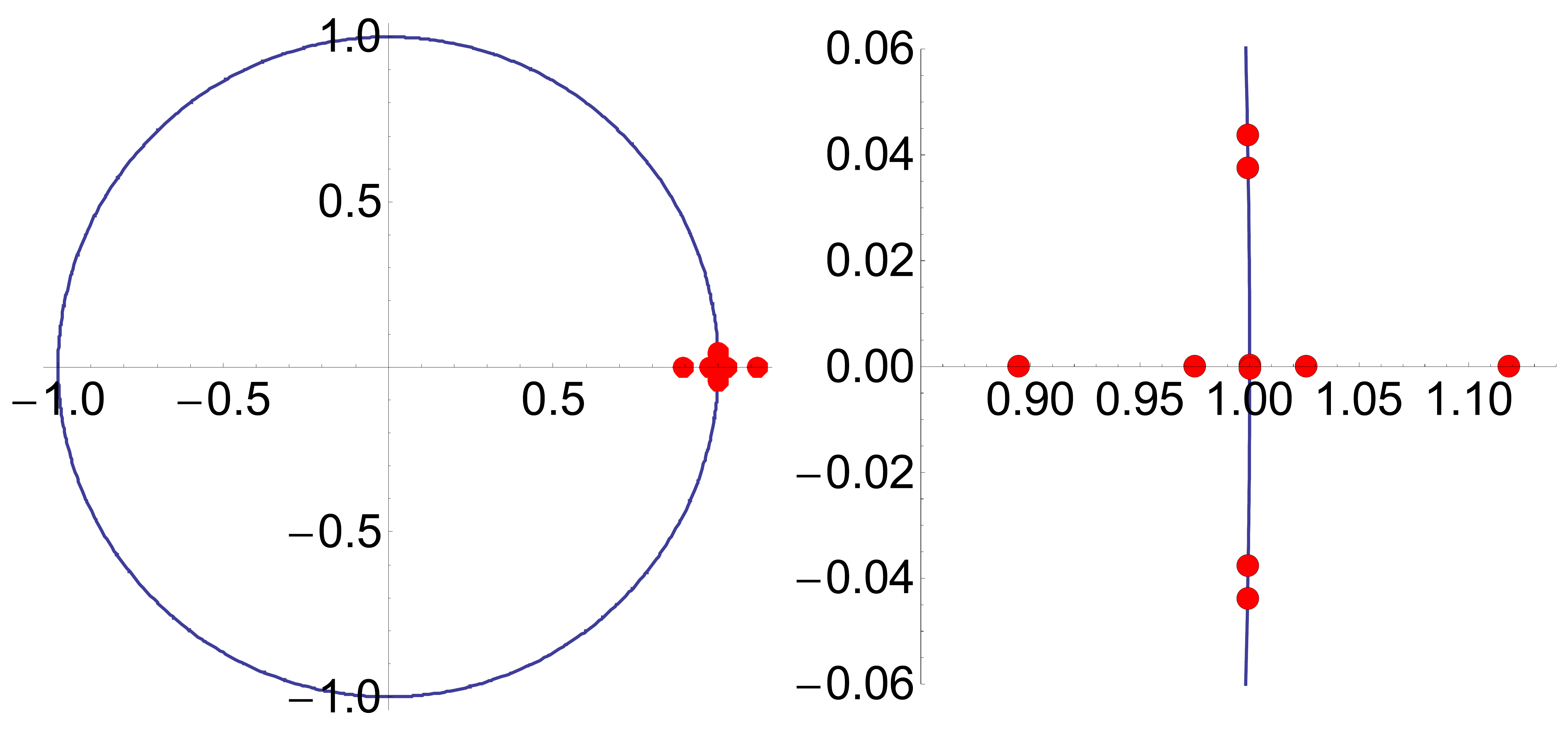}\hspace{-0.5cm}{\rm (b)}\vspace{-0.25cm}\\
\includegraphics[width=0.472\textwidth]{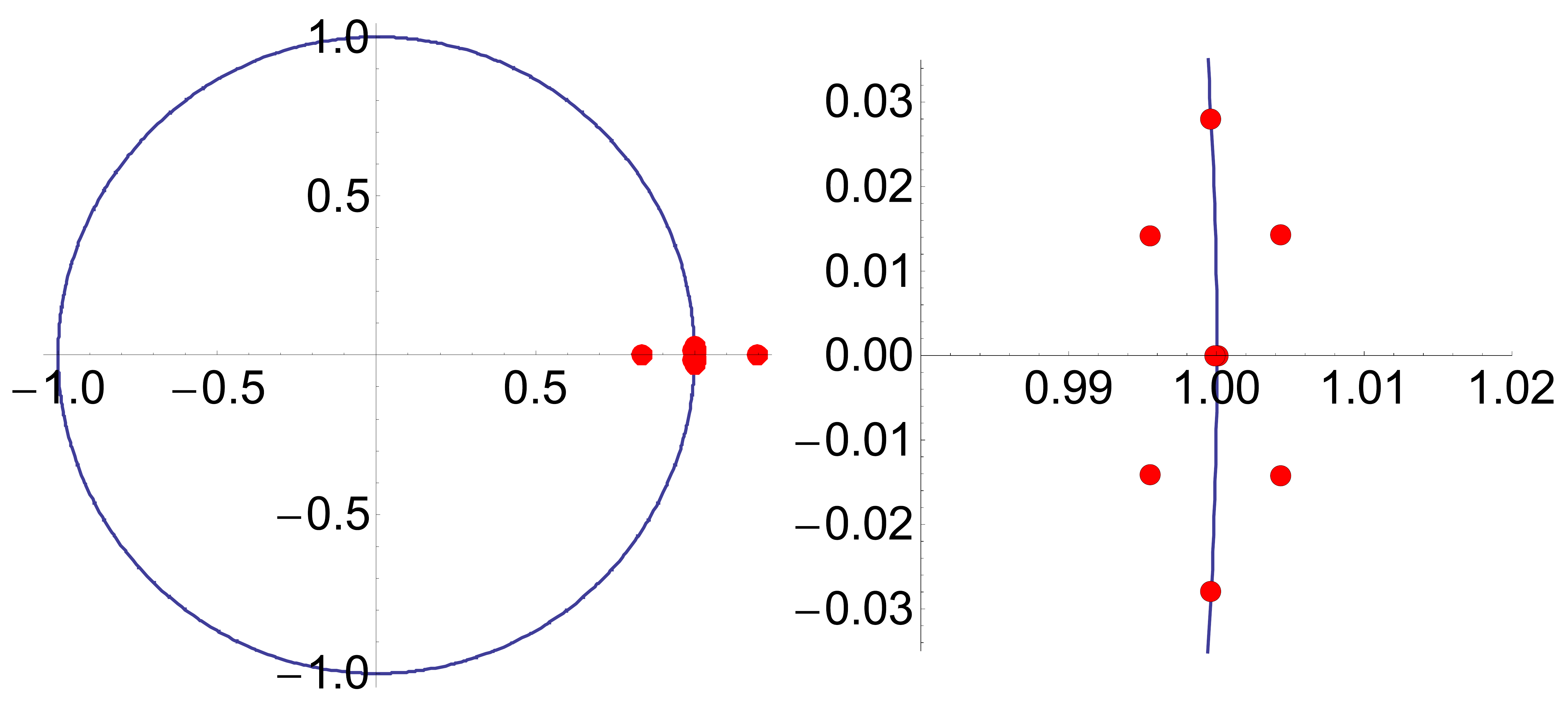}\hspace{-0.5cm}{\rm (c)}\vspace{-0.25cm}\\
\includegraphics[width=0.472\textwidth]{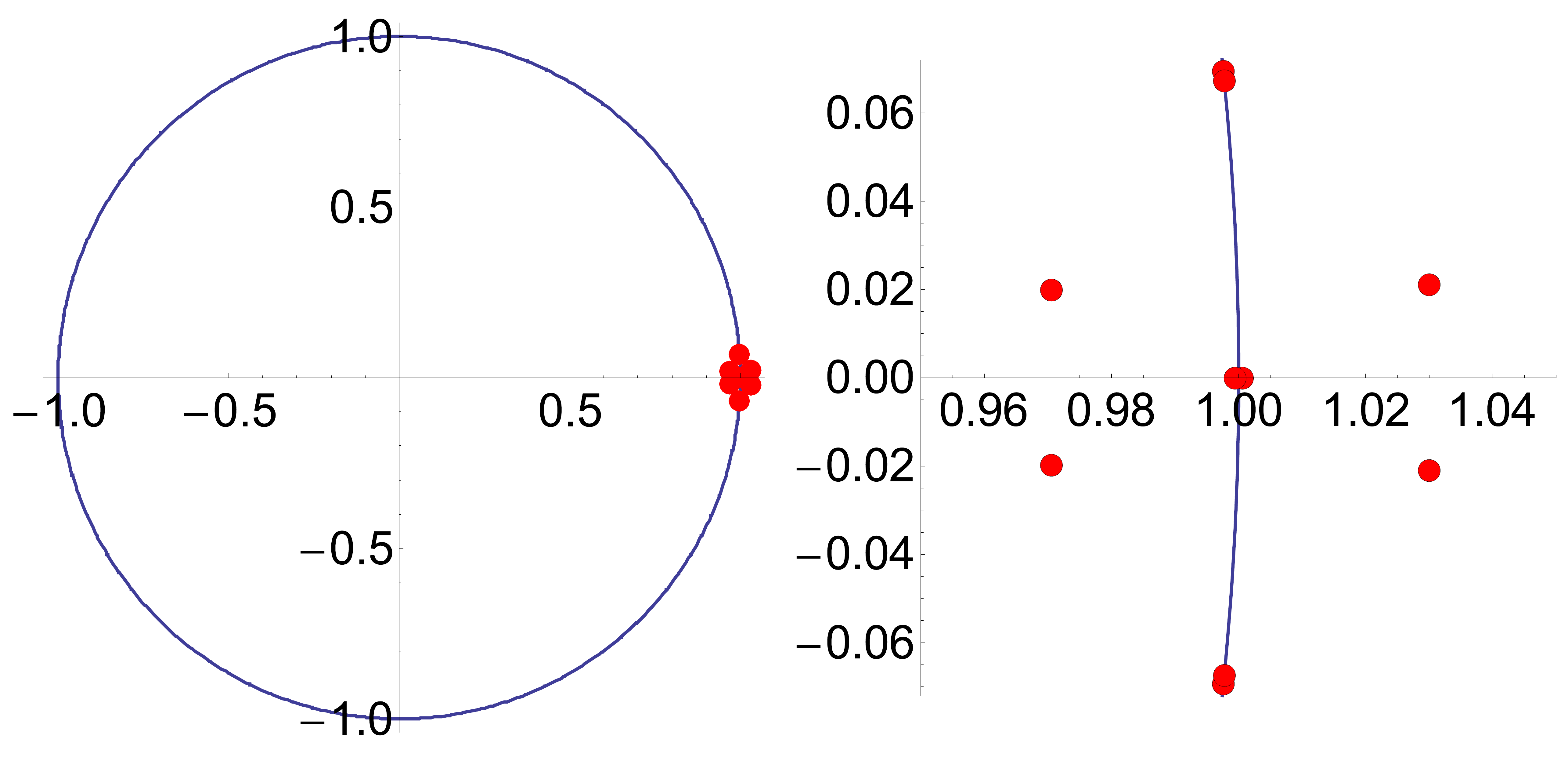}\hspace{-0.5cm}{\rm (d)}\vspace{-0.25cm}\\
\includegraphics[width=0.472\textwidth]{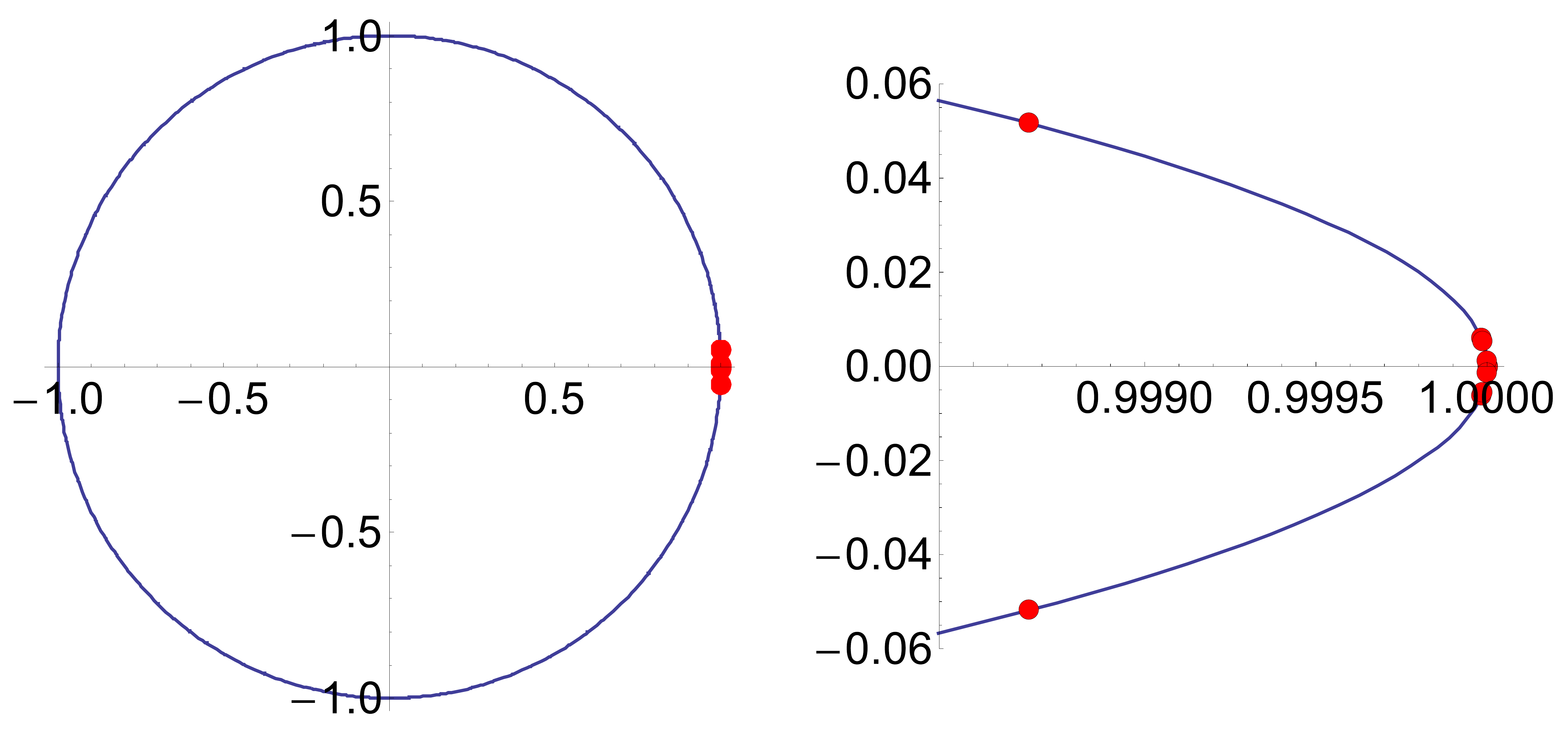}\hspace{-0.5cm}{\rm (e)}\vspace{-0.28cm}\\
\end{array}$
\caption{Examples of distribution of eigenvalues (red dots) with respect to the complex unit circle with a magnification on the right of each panel for particular orbits from the families of the 1:2:3 resonant chain. (a) Triple instability with three pairs of real eigenvalues ($S_1$ family); (b) Double instability with two pairs of real eigenvalues ($S_3$ family); (c) u-complex instability with one pair of real and two pairs of complex eigenvalues outside the unit circle ($S_5$ family); (d) Complex instability with two pairs of complex eigenvalues outside the unit circle, while the rest remain on it ($S_5$ family). (e) Linear stability with all pairs on the unit circle ($S_1$ family).}\label{eigen}
\end{figure}

\section{Orbital stability}\label{stab}

\subsection{Linear stability of periodic orbits}\label{linear}

Let us denote with the vector $\mathbf{x}=(x_1,...,x_{10})$ the set of ten variables of the system $\left\{x_1,x_2,x_3,y_2,y_3,\dot x_1,\dot x_2, \dot x_3, \dot y_2, \dot y_3\right\}$. Then the solution $\mathbf{x}=\mathbf{x}(t;\mathbf{x}_0)$  corresponds to the initial conditions $\mathbf{x}_0=\mathbf{x}(0)$. The variational equations of the system in Eq. \ref{eqn:all-lines} and their solutions are written as:

\begin{equation}
\begin{array}{ccc}
\dot{\bm{\eta}}=\textbf{J}(t)\bm{\eta} & \Rightarrow & \bm{\eta}=\Delta(t)\bm{\eta}_0, 
\end{array}
\end{equation}
where \textbf{J} is the Jacobian matrix of the system and $\Delta (t)$ is the fundamental matrix of solutions (called also matrizant or state transition matrix). 

If the solution $\mathbf{x}(t;\mathbf{x}_0)$ corresponds to a periodic orbit of period $T$, $\Delta (T)$ is the monodromy matrix. If and only if all the eigenvalues (shown with red dots in Fig. \ref{eigen}) of $\Delta (T)$ lie on the complex unit circle, the periodic orbit is classified as linearly stable and $\bm{\eta}(t)$ remains bounded. 

We remark that the eigenvalues are in conjugate pairs, as $\Delta (T)$ is symplectic. Moreover, due to the existence of the energy integral, one pair of eigenvalues (denoted here by $\lambda_1$ and $\lambda_2$) is always equal to unity. Some possible configurations of the other four pairs of eigenvalues and different types of instability are shown in Fig. \ref{eigen}.

\subsection{Chaotic Indicator}\label{chaos}

\begin{figure}[h!]%[ht!]
\begin{center}
\includegraphics[width=0.48\textwidth]{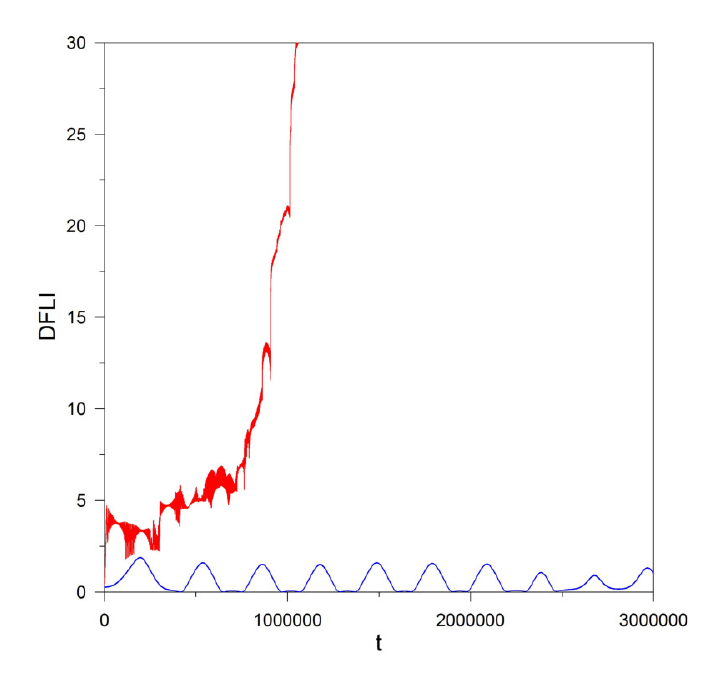}%\vspace{-0.4cm}\\
\end{center}
\caption{Evolution of DFLI for an unstable (red) and a linearly stable (blue) periodic orbit with eigenvalues shown in panel (a) and (e) in Fig. \ref{eigen}, respectively.}\label{fdfli}
\end{figure}

Establishing whether the eigenvalues lie on the unit circle or not can sometimes become an ambiguous process due to the limited accuracy. 
Therefore, apart from the linear stability along a family of periodic orbits, we also computed a chaotic indicator and, in particular, a simple detrended Fast Lyapunov Indicator \citep[DFLI, e.g.,][]{voyatzis08} defined as 
\begin{equation}\label{dfli}
DFLI(t)=log \left ( \frac{1}{t}||{\bm{\xi}}(t)|| \right ),\vspace{-0.5cm}
\end{equation}
where $\bm{\xi}$ is the deviation vector computed after numerical integration of the variational equations. In Fig. \ref{fdfli}, we illustrate the DFLI's behavior for the system Kepler-51. For a stable periodic orbit its evolution remains almost constant over time and takes small values, whereas it increases exponentially when the orbit is unstable.   

\end{appendix}

\end{document}